\documentclass{revtex4}
\usepackage{graphicx}
\usepackage{bm}
\usepackage{amsmath}
\usepackage[dvipsnames]{color}

\usepackage{pstricks}
\usepackage{epsfig}
\usepackage{amsbsy}
\usepackage{varioref}
\usepackage{pifont}
\usepackage{axodraw}

\def\slashii#1{\setbox0=\hbox{$#1$}             
   \dimen0=\wd0                                 
   \setbox1=\hbox{\sl/} \dimen1=\wd1            
   \ifdim\dimen0>\dimen1                        
      \rlap{\hbox to \dimen0{\hfil\sl/\hfil}}   
      #1                                        
   \else                                        
      \rlap{\hbox to \dimen1{\hfil$#1$\hfil}}   
      \hbox{\sl/}                               
   \fi}                                         %
\def\slashiii#1{\setbox0=\hbox{$#1$}#1\hskip-\wd0\hbox to\wd0{\hss\sl/\/\hss}}
\newcommand{\beq}{\begin{equation}}
\newcommand{\eeq}{\end{equation}}
\newcommand{\bq}{\begin{equation}}
\newcommand{\eq}{\end{equation}}
\newcommand{\ba}{\begin{array}}
\newcommand{\ea}{\end{array}}
\newcommand{\beqa}{\begin{eqnarray}}
\newcommand{\eeqa}{\end{eqnarray}}
\newcommand{\infinity}{\infty}

\def\dif{\partial}

\def\TT{{T}}
\def\T{{T}}
\def\Tb{\overline{T}}

\def\Tt{\widetilde{T}}
\def\Tb{\overline{T}}
\def\Tbt{\widetilde{\overline{T}}}

\def\End{\end{document}}

\def\to{\rightarrow}

\def\dis{\displaystyle}
\def\f{\frac}
\def\ov{\overline}
\def\[{\left[}
\def\]{\right]}
\def\({\left(}
\def\){\right)}

\def\pit{\widetilde{\pi}}

\def\U1EM{U(1)_{\rm em}}

\def\T{{T}}

\def\geqq{\geq}

\def\gb{\overline{g}}

\def\[{\left[}
\def\]{\right]}
\def\dis{\displaystyle}

\def\si{\sigma}

\def\d{\delta}

\def\hf{\frac{1}{2}}

\def\MM{\mathfrak{M}}

\def\Ab{\mathbf A}

\def\ep{\epsilon}

\def\RRB{\mathbf{R}}
\def\RRBT{\widetilde{\mathbf{R}}}
\def\AAB{\mathbf{A}}
\def\WWB{\mathbf{W}}

\def\CF{\mathcal F}
\def\CG{\mathcal G}
\def\CD{\mathcal D}

\def\QQB{\mathbf{Q}}
\def\MMB{\mathbf{M}}
\def\MMW{\mathbf{M}_{W}^{\,2}}

\def\MMWT{\widetilde{\mathbf{M}}_{W}^{\,2}}

\def\MM{\mathfrak{M}}
\def\MMBT{\widetilde{\mathbf{M}}}

\def\Pit{\widetilde{\Pi}}
\def\pit{\widetilde{\pi}}

\def\RRB{\mathbf{R}}
\def\RRBT{\widetilde{\mathbf{R}}}
\def\AAB{\mathbf{A}}
\def\WWB{\mathbf{W}}

\def\CF{\mathcal F}
\def\CG{\mathcal G}
\def\CD{\mathcal D}

\def\QQB{\mathbf{Q}}
\def\MMB{\mathbf{M}}
\def\MM{\mathfrak{M}}
\def\MMBT{\widetilde{\mathbf{M}}}

\def\Pit{\widetilde{\Pi}}
\def\pit{\widetilde{\pi}}

\def\Eb{\overline{E}}
\def\Eh{\widehat{E}}
\def\Mb{\overline{M}}
\def\Mh{\widehat{M}}
\def\eph{\widehat{\epsilon}}
\def\pib{\mbox {\boldmath$\pi$}}
\def\Ab{\mathbf A}

\def\ep{\epsilon}

\def\Gt{\widetilde{G}}

\def\gz{g_0^{~}}
\def\gzz{g_0^{2}}
\def\gb{g_1^{~}}
\def\gbb{g_1^{2}}

\def\dif{\partial}

\newcommand{\bbM}{\MMB}
\newcommand{\bbQ}{\QQB}
\newcommand{\bbR}{\RRB}

\begin{document}
\title{General Sum Rules for $\boldsymbol{WW}$ Scattering in Higgsless Models: \\ Equivalence Theorem and Deconstruction Identities}
\author{R. Sekhar Chivukula$^a$}
\email[email: ]{sekhar@msu.edu}
\author{Hong-Jian He$^b$}
\email[email:]{hjhe@mail.tsinghua.edu.cn}
\author{Masafumi Kurachi$^c$}
\email[email:]{kurachi@yukawa.kyoto-u.ac.jp}
\author{Elizabeth H. Simmons$^a$}
\email[email: ]{esimmons@msu.edu}
\author{Masaharu Tanabashi$^d$}
\email[email:]{tanabash@eken.phys.nagoya-u.ac.jp}
\vspace{.25cm}
\affiliation{$^a$Department of Physics and Astronomy\\ Michigan State University\\ East Lansing, MI 48824\\ }
\affiliation{$^b$Center for High Energy Physics\\ Tsinghua University\\ Beijing 100084, China\\ }
\affiliation{$^c$Yukawa Institute for Theoretical Physics\\Kyoto University\\Kyoto 606-8502, Japan\\}
\affiliation{$^d$Department of Physics\\ Nagoya University\\ Chikusa-ku Nagoya 464-8602  Japan}

\date{\today}

\begin{abstract}
We analyze inelastic $2\to 2$ scattering amplitudes for 
 gauge bosons and Nambu-Goldstone bosons in deconstructed Higgsless models.  Using the (KK) Equivalence Theorem in 4D (5D), we derive a set of general sum rules among the boson masses and multi-boson couplings that are valid for arbitrary deconstructed models.  
Taking the continuum limit, our results naturally include the 5D Higgsless model sum rules 
 for arbitrary 5D geometry and boundary conditions; they also reduce to the elastic sum rules when applied to the special case of elastic scattering. For the case of linear deconstructed Higgsless models, we demonstrate that the sum rules can also be derived from a set of general deconstruction identities and completeness relations.  We apply these sum rules to the deconstructed 3-site Higgsless model and its extensions; we show that in 5D ignoring all higher KK modes ($n>1$) is inconsistent once the inelastic channels become important.  Finally, we discuss how our results generalize beyond the case of linear Higgsless models.

\end{abstract}

\maketitle

\section{Introduction}

Higgsless models \cite{Csaki:2003dt}  break the electroweak symmetry without invoking a fundamental scalar  Higgs boson \cite{Higgs:1964ia}.  These models were originally motivated by ideas of  gauge/gravity duality \cite{Maldacena:1998re,Gubser:1998bc,Witten:1998qj,Aharony:1999ti}, and are in a sense ``dual'' to more conventional models of dynamical symmetry breaking \cite{Weinberg:1979bn,Susskind:1979ms} such
as ``walking techicolor'' \cite{Holdom:1981rm,Holdom:1985sk,Yamawaki:1986zg,Appelquist:1986an,Appelquist:1987tr,Appelquist:1987fc}.  In the absence of a Higgs boson in the model, some other state must take on the role of unitarizing  the scattering of longitudinally polarized electroweak gauge bosons; the extra vector bosons present in Higgsless models play this role \cite{SekharChivukula:2001hz,Chivukula:2002ej,Chivukula:2003kq,He:2004zr}. 
Higgsless models provide effectively unitary descriptions of the electroweak sector beyond the TeV energy scale.  On their own, Higgsless models are not renormalizable because they are based on the physics of  TeV-scale \cite{Antoniadis:1990ew} compactified $SU(2)^2 \times U(1)$ five-dimensional gauge theories (with appropriate boundary conditions) \cite{Agashe:2003zs,Csaki:2003zu,Burdman:2003ya,Cacciapaglia:2004jz}.  Instead, Higgsless models are properly viewed as effective field theories valid  below a cutoff energy scale inversely proportional to the five-dimensional gauge-coupling squared. Above this energy scale, a new ``high-energy" completion, which is valid to higher energies, must enter to describe the physics appropriately.   

\begin{figure}[hb]
\label{fig:Fig1}
\centering
\includegraphics[width=0.6\textwidth]{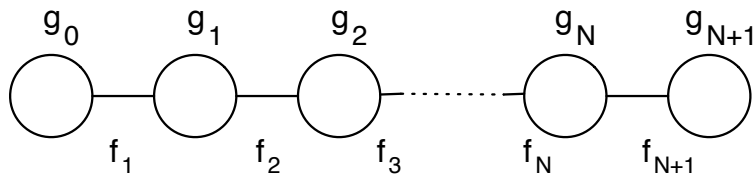}
\caption
{A deconstructed model, in moose notation \protect\cite{Georgi:1986hf}, corresponding
to an arbitrary five-dimensional gauge theory. The coupling constants ($g_i$)
and $f$-constants ($f_j$) are arbitrary, corresponding, as discussed in the text, 
 to the position-dependent coupling and warp factors chosen.}
\end{figure}

Since a compactified five-dimensional
theory gives rise to an infinite tower of massive four-dimensional ``Kaluza-Klein" (KK) modes of the 
gauge field, one might anticipate that a low-energy approximation would involve only a finite number of low-mass states. Deconstruction 
\cite{Arkani-Hamed:2001ca,Hill:2000mu} is a technique that realizes this 
expectation. As illustrated in Fig.\ref{fig:Fig1}, deconstruction
may be interpreted as replacing the continuous fifth-dimension with a 
lattice of individual gauge groups (the factors of the semi-simple  
deconstructed four-dimensional gauge group) at separate sites in ``theory space". 
The ``link" variables connecting the gauge groups
at adjacent sites represent symmetry breaking fields which break the groups at adjacent
sites down to the diagonal  subgroup, as conventional
in ``moose notation" \cite{Georgi:1986hf}. In the continuum limit (in which the number of
sites on the deconstructed lattice is taken to infinity), the kinetic energy terms for the link
variables give rise to the
terms in the five-dimensional kinetic energy involving derivatives with respect to
the compactified coordinate.
Deconstructed Higgsless models \cite{Foadi:2003xa,Hirn:2004ze,Casalbuoni:2004id,Chivukula:2004pk,Perelstein:2004sc,Georgi:2004iy,SekharChivukula:2004mu} have been used as tools to compute the general properties of Higgsless theories, and to illustrate the phenomenological properties of this class of models \footnote{The gauge-structure of deconstucted Higgsless models with only a few extra vector bosons  are equivalent to models of extended electroweak gauge symmetries
\protect\cite{Casalbuoni:1985kq,Casalbuoni:1996qt}  motivated by models of hidden local symmetry \protect\cite{Bando:1985ej,Bando:1985rf,Bando:1987br,Bando:1988ym,Bando:1988br,Harada:2003jx}.}.

Recently, we studied elastic pion-pion scattering in global linear moose models \cite{SekharChivukula:2006we} and showed that the Kadanoff-Wilson block spin transformation \cite{Kadanoff:1966wm,Wilson:1971bg} and the Equivalence Theorem (ET) \footnote{For a generalization
of the Equivalence Theorem to compactified five-dimensional theories, see 
\protect{\cite{SekharChivukula:2001hz}}.} \cite{Cornwall:1974km,Lee:1977eg,Chanowitz:1985hj,Yao:1988aj,Bagger:1990fc,He:1992ng,He:1994yd,He:1997cm} render our analysis applicable to $W_L W_L$ elastic scattering in Higgsless models with arbitrary background 5D geometry, spatially dependent gauge-couplings, and brane kinetic energy terms for the gauge-bosons.  The form of the high-energy elastic scattering amplitudes in Higgsless models is particularly 
interesting because it can potentially yield an upper bound on the scale at which the 
effective field theory becomes strongly coupled.  We concluded that elastic unitarity provides a useful guide to the range of a model's validity only for models in which the lightest KK mode has a mass greater than about 700 GeV; otherwise, inelastic channels such as  $\pi\pi \to W_1 W_1$ become available and important.  

Our aim in this paper is to derive general identities and sum rules that apply to the masses and couplings of the gauge and Nambu-Goldstone KK modes in $WW$ scattering in an
arbitrary deconstructed Higgsless model, including inelastic scattering channels. We derive these sum rules by application of the KK equivalence theorem, and therefore we see that these sum rules are a result of the gauge-invariance of the theory.   
Taking the continuum limit, our results naturally include all 5D Higgsless model
sum rules  for arbitrary 5D geometry and boundary conditions, and they also reduce to the elastic sum rules \cite{Csaki:2003dt} when applied to the special case of elastic scattering. For the case of deconstructed Higgsless models corresponding to linear mooses \cite{Georgi:1986hf}, we will demonstrate that the sum rules can also be derived from a set of general deconstruction identities and completeness relations.  We apply these sum rules to the deconstructed 3-site Higgsless model and its extensions; we show that in 5D ignoring all higher KK modes ($n>1$) is inconsistent once the inelastic channels become important.  Finally, we discuss how our results generalize beyond the case of linear Higgsless models.

In section II, we review the features of Higgsless models that are essential for our analysis: the Lagrangian, principles of deconstruction, the shift from gauge to mass eigenbasis for both the vector bosons and the ``eaten" Nambu-Goldstone bosons present in the $R_\xi$ gauge, and the definitions of the couplings among various states in the spectrum.  Beginning in section II, we simplify our
analysis by considering the case of an $SU(2)^{N+1}$ deconstructed model, in which the
gauge symmetries are completely broken -- corresponding in the continuum limit to
a Higgsless model of $SU(2)_W$ breaking without hypercharge (and therefore without
a massless photon). The generalization to the phenomenologically
realistic case of $SU(2)^{N+1} \times U(1)^{M}$ \cite{SekharChivukula:2004mu} is straightforward, and is discussed in 
section VII. 

In sections III and IV we use two-body scattering of $W$ bosons and of Nambu-Goldstone bosons to derive general sum rule relationships among the masses and couplings of the KK mass eigenstates.  First, we present our results for the leading contributions (in power of energy) to these processes.  Then we apply the KK Equivalence Theorem to those scattering processes, which yields the sum rules listed in Table \ref{tab:sumrules}; an Appendix shows how the sum rules are realized in the specific case of the three-site Higgsless model \cite{SekharChivukula:2006cg}.  Our analysis considers inelastic scattering, which allows us to derive a wider class of sum rules; we also show how the sum rules simplify in the elastic limit and compare with the specific 5D elastic results in the literature \cite{Csaki:2003dt}.  Similarly, we derive the sum rules in 4D deconstructed models but also show how the sum rules appear in the continuum limit (and again compare with the special case in \cite{Csaki:2003dt}).  The identities derived in Section IV apply to the multi-boson couplings and masses for any consistent gauge theory in 4D or 5D;  while we take a linear deconstructed model as our benchmark, the form of the relations transcends this.  

We analyze the implications of gauge invariance in section V, by considering the 2-body
$WW$-amplitude and Nambu-Goldstone $\widetilde{\pi}\widetilde{\pi}$-amplitude in
$R_\xi$ gauge. We derive  a new condition on the masses and couplings, and show that the 
sum rules obtained from the ET  in section IV  hold in $R_\xi$ gauge.

Section VI presents an alternative way of deriving relations among the couplings and boson masses in Higgsless models.  We start by deriving completeness relations based explicitly on the orthogonality of the KK gauge-boson (and KK Nambu-Goldstone-boson) mass-eigenstate wavefunctions in the deconstructed bulk.  Then, we derive a set of Ward-Takahashi identities relating each coupling involving Nambu-Goldstone modes to couplings of gauge modes only; these are listed in Table \ref{tab:wtcr}.  
Some of these results are specific to linear moose models and their 5D continuum limits because their derivation draws directly upon the specific form of the mixed gauge/Nambu-Goldstone couplings.  Within this class of models, however, the deconstruction identities derived in Section VI form a generalization of the ET sum rules obtained in sections IV and V because they are not restricted to $nn\to mm$ scattering processes and allow all of the coupling indices to vary independently.  We demonstrate the relationships between the deconstruction identities and the ET sum rules of Section IV; these
relationships are summarized in Table \ref{tab:sumrules}.

Section VII summarizes the relation between the two derivations and the range of applicability of each set of results.  We also demonstrate how our sum rules generalize some specific elastic results in \cite{Csaki:2003dt}.  In addition, we use the sum rules to show that in 5D ignoring all higher KK modes ($n>1$) is inconsistent once inelastic scattering channels become important, in contrast to the scenario adopted by Ref. \cite{Birkedal:2004au}.  Finally, we indicate how the results in the present paper will be used in a forthcoming work to make a more accurate assessment of the scale at which two-body scattering (including inelastic coupled channels) violates unitarity.

%
\begin{table}[hbtp]
\centering
\begin{tabular}{|c|c|c|c||c||c|}
    \hline
    Derived&At & Eq.& {\bf Equivalence Theorem Sum Rule for Inelastic} & Elastic& Related to the following \\
      From the&Order&\#& {\bf $nn \to mm$ Scattering with $(n\neq m)$} & (n=m) & Completeness Relations (CR)\\
      Processes&&&& Eq. \#& \& Ward-Takashi Identities (WTI)\\
    \hline\hline
&&&&&\\
 $LL\to LL$&$E^4$&(\protect\ref{eq:E4-sum1}) & $G_4^{nnmm} = \sum_k G_3^{nnk}G_3^{mmk}$ &   (\protect\ref{eq:E4-sum-el})& Both are special cases of  \\
 $LL\to LT$ & $E^3$&&&& \\
 $LL\to TT$ & $E^2$ &(\protect\ref{eq:E4-sum2}) & $G_4^{nnmm} = \sum_k \(G_3^{nmk}\)^2$ & (\protect\ref{eq:E4-sum-el})& gauge-boson CR (\ref{eq:id1})\\
 $LT \to TT$ & $E^1$&&&&\\
    \hline\hline
&&&&&\\
$LL\to LL$ &$E^0$&  (\protect\ref{eq:E0-sum2}) & $ G_4^{nnmm}M_-^4 \!+\! \sum_k\!\[\(\!G_3^{nmk}\!\)^{\!2}\!\!
 -\! G_3^{nnk}G_3^{mmk}\]\!\!\(M_k^4\!\!-\! 2M_+^2M_k^2\)$ &&Both are special cases of \\
&  && = $M_n^2M_m^2\!\sum_k\!\!\[\!\(\!\Gt_3^{nmk}\!\)^{\!2}\!\!
 -\! \Gt_3^{nnk}\Gt_3^{mmk}\!\]\! $ & N/A& gauge-boson CR (\protect\ref{eq:id1}) \\
$LL \to TT$ & $E^0$ &(\protect\ref{eq:SR-E0-LLTT-2in})&$ G_4^{nnmm} =
 \sum_k\hf\!\[  \Gt_3^{nnk}G_3^{mmk}
                 +G_3^{nnk}G_3^{mmk}\Mb_k^2 \]$ & (\protect\ref{eq:SR-E0-LLTT-2e})&and  $\tilde{G}_3$ WTI (\protect\ref{eq:WT-t3}) \\
                 &&&&&\\
\hline\hline
&&&&&\\
$LL\to LL$ &$E^2$&(\protect\ref{eq:E2-sum2})& $\sum_k\[\(\!G_3^{nmk}\!\)^2 - G_3^{nnk}G_3^{mmk}\]\! M_k^2
= \sum_k\(\!G_3^{nmk}\!\)^2\f{M_-^4}{M_k^2}$& N/A&NG-boson CR (\protect\ref{eq:id2}) and\\ 
&&&&& $\tilde{G}_{31}$ WTI (\protect\ref{eq:WT-t31})\\
\hline\hline
&&&&&\\
$LL\to LL$ &$E^2$ & (\protect\ref{eq:E2-sum1}) & $2M_+^2G_4^{nnmm} + \sum_k\(\!G_3^{nmk}\!\)^2
\[\f{M_-^4}{M_k^2} - 3M_k^2\] = \f{4M_n^2M_m^2}{v^2}\Gt_4^{nnmm}$& (\protect\ref{eq:E2-sum-el})  & NG-boson CR (\protect\ref{eq:id2}) and\\
&&&&& $\tilde{G}_{31}$ and $\tilde{G}_4$ WTI's (\protect\ref{eq:WT-t31}, \protect\ref{eq:WT-t4})\\
\hline\hline
&&&&&\\
 $LL\to LL$ &$E^0$&(\protect\ref{eq:E0-sum1}) & $\sum_k\!\(\!G_3^{nmk}\!\)^{\!2}\!
 \[\f{\,M_k^2-M_+^2\,}{M_nM_m}\]^2
 =  \sum_k\(\!\Gt_3^{nmk}\!\)^2$ & (\protect\ref{eq:E0-sum1-el}) & Both hold term by term  \\
 &&&&&for each $k$ because of\\
$LL \to TT$ & $E^0$ & (\protect\ref{eq:SR-E0-LLTT-1in})&  $\sum_k (\Gt_3^{nkm})^2 =
 \sum_k (G_3^{nmk})^2\!\(\!\Mb_k - \f{\Mb_-^2}{\Mb_k}\!\)^{\!\!2}$ & (\protect\ref{eq:SR-E0-LLTT-1e}) &  $\tilde{G}_3$ WTI (\protect\ref{eq:WT-t3}) \\
 \hline\hline
&&&&&\\
 $LL\to LT$&$E^1$&(\protect\ref{eq:G41a}) &  $\Gt_{41}^{nnmm} = -\f{2M_m}{v}\Gt_4^{nnmm}$ & (\protect\ref{eq:G41a}) & $\tilde{G}_{41}$ and $\tilde{G}_4$ WTI's (\protect\ref{eq:WT-t41},\protect\ref{eq:WT-t4}) \\
 &&&&&\\
 \hline\hline
 &&&&&\\$LL \to TT$ & $E^0$ &(\protect\ref{eq:SR-E0-LLTT-3in})&$\Gt_{42}^{nnmm} = G_4^{nnmm}\Mb_m^2 -\sum_k(G_3^{nmk})^2\Mb_k^2 $ &(\protect\ref{eq:SR-E0-LLTT-3e})& $\tilde{G}_{42}$ WTI (\protect\ref{eq:WT-t42})\\ 
 &&&&&\\
\hline\hline\hline
   && & {\bf Gauge Invariance Relation for  Inelastic} & & \\
      &&& {\bf $nn \to mm$ Scattering with $(n\neq m)$} &  &\\
 \hline\hline
 All &&&&&\\
polarizations & N/A & (\protect\ref{eq:xi-cond-G3x})& $(\Gt_{31}^{nmk} M_k)^2 = (G_3^{nmk} (M_n - M_m))^2 $& N/A & $\tilde{G}_{31}$ WTI (\protect\ref{eq:WT-t31})\\
  \hline\hline 

  \end{tabular}
  \caption{Higgsless model sum rules derived in this paper.  The broad central column lists each sum rule derived from applying the Equivalence Theorem (ET) to the inelastic ($n\neq m$) scattering processes involving longitudinal gauge bosons in Section \protect\ref{ETsumsec} or gauge invariance in Section V.  To the left of each sum rule is its equation number in the text; to the right of each sum rule with an elastic ($n=m$) version is the corresponding text equation number.  The two left-most columns of the table indicate from which process and at what order in powers of energy a given sum rule has been derived.  The right-most column of the table summarizes how each sum rule is related to the completeness relations and deconstruction identities derived for Higgsless models in Section \protect\ref{WTI-CR-sec}. Note that the continuum 5D versions of these identities are obtained by setting both $\tilde{G}_4$ and $\tilde{G}_{41}$ to zero.}
  \label{tab:sumrules}
\end{table}

%
\begin{table}[hbtp]
\centering
\begin{tabular}{|c|c|c|}
    \hline
    Description& Completeness Relations (CR) \ \& \ Ward-Takahashi Identities (WTI) & Eq. \#\\
    \hline\hline
&&\\
gauge-boson CR& $\sum_i G_{3}^{nmi} G_{3}^{\ell ki}
  = G_4^{nm\ell k}$&(\protect\ref{eq:id1})\\
\hline
&&\\
NG-boson CR &$ \sum_i (M_n + M_m)\tilde{G}_{31}^{nmi}
         (M_\ell + M_k)\tilde{G}_{31}^{\ell ki}
  =   
  \sum_i (G_3^{nki} G_3^{m\ell i} - G_3^{n\ell i} G_3^{mk i})M_i^2$&(\protect\ref{eq:id2})\\
\hline\hline
&&\\
$\tilde{G}_{31}$ WTI &$  \tilde{G}_{31}^{nmk} M_k 
  = - G_3^{nmk} (M_n - M_m)$&(\protect\ref{eq:WT-t31})\\
\hline
&&\\
$\tilde{G}_{3}$ WTI&$\tilde{G}_{3}^{nmk} M_n M_m 
  = G_3^{nmk}(M_n^2 + M_m^2 - M_k^2)$&(\protect\ref{eq:WT-t3})\\
\hline
&&\\
$\tilde{G}_{42}$ WTI&~~~~$\tilde{G}_{42}^{nm\ell k} M_n M_m
  = \frac{1}{2} G_4^{nm\ell k}(M_\ell^2 + M_k^2)
     -\frac{1}{2} \sum_i 
     (G_3^{n\ell i} G_3^{mki} + G_3^{nki}G_3^{m\ell i})M_i^2$~~~~&(\protect\ref{eq:WT-t42})\\
\hline
&&\\
$\tilde{G}_{41}$ WTI& $\tilde{G}_{41}^{nm\ell k} M_n M_m M_\ell
  =-\frac{v}{2} G_4^{nm\ell k}(M_n^2 + M_m^2 + M_\ell^2 + M_k^2)$&(\protect\ref{eq:WT-t41})\\
  &$\ \ \ \ \ \ \ \ \ \ \ \ \ \ \ \ \ \ \ \ \ \ \ +\frac{v}{2} \sum_i 
      (G_3^{nmi} G_3^{\ell k i} 
      +G_3^{n\ell i} G_3^{m k i}
      +G_3^{nki} G_3^{m \ell i}) M_i^2$&\\
\hline
&&\\
$\tilde{G}_{4}$ WTI&$ \tilde{G}_{4}^{nm\ell k} M_n M_m M_\ell M_k
  =
     \dfrac{v^2}{4} 
     G_4^{nm\ell k}(M_n^2 + M_m^2 + M_\ell^2 + M_k^2)$&(\protect\ref{eq:WT-t4})\\
     &$\ \ \ \ \ \ \ \ \ \ \ \ \ \ \ \ \ \ \ \ \ \ \     -\dfrac{v^2}{4} \sum_i 
      (G_3^{nmi} G_3^{\ell k i} 
      +G_3^{n\ell i} G_3^{m k i}
      +G_3^{nki} G_3^{m \ell i}) M_i^2$&\\
 \hline\hline

  \end{tabular}
  \caption{Higgsless model completeness relations and Ward-Takahashi identities derived in this paper.  The broad central column lists each completeness relation or WT identity.  To the left of each equation is a brief description; to the right is the equation number in the text of Section \protect\ref{WTI-CR-sec}.}
  \label{tab:wtcr}
\end{table}


\section{Higgsless Models in General}
\label{sec:higgsless}

\begin{figure}
\label{fig:Fig2}
\centering
\includegraphics[width=0.7\textwidth]{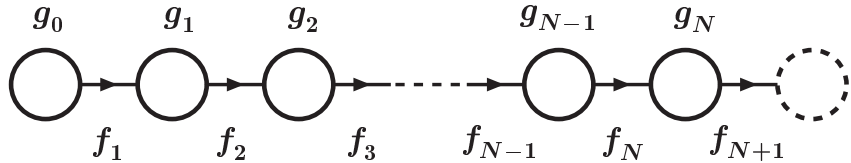}
\vspace*{-9mm}
\caption{The linear moose for which we study $WW$ scattering in this paper.  Site 
$N+1$ is a global $SU(2)$ group ($g_{N+1}=0$) and sites $0$ to $N$ are $SU(2)$ gauge groups.
This corresponds to the deconstructed form of either flat space or AdS 5-dimensional models of 
electroweak symmetry breaking depending on the couplings
and $f$-constants chosen.
}
\end{figure}

In this section we review the essential elements of deconstructed linear Higgsless models and  their 5D counterparts. Because our analysis will focus on high energy $WW$-scattering, we can
replace the $U(1)$ gauge group at site $N+1$ by a global $SU(2)$ group, as shown in the
semi-open moose of Fig.\ref{fig:Fig2}; the 5-d continuum limit of this moose contains theories of the type 
 \,$SU(2)_{\rm 5D}^p$\, ($p=1,2,\cdots$).  Our analysis will leave arbitrary the values of the gauge couplings and $f$-constants.
In the continuum limit, therefore, our results apply to models with arbitrary 5D background geometry,
spatially dependent gauge-couplings, and brane kinetic energy terms for the gauge-bosons. 
 
The Lagrangian of the deconstruction in Fig.\ref{fig:Fig2} is,
\begin{equation}
\label{eq:L}
\dis
  {\cal L} \,~=~\,
  \sum_{j=0}^{N} -\dfrac{1}{\,2\,}
                   \mbox{Tr}\(F_{j\mu\nu} F_j^{\mu\nu}\)
~+\,
  \sum_{j=1}^{N+1} \dfrac{\,f_j^2\,}{4} \mbox{Tr}\left[ (D_\mu U_j)^\dagger (D^\mu U_j) \right],
\end{equation}
where the gauge fields on each site are represented by \,$\Ab_j^\mu\equiv A_j^{a\mu}T^a \in SU(2)_j$\,
($j=0,\cdots,N$), the links are non-linear sigma model fields $U_j = \exp\[2i\pib_j^{~}/f_j^{~}\] \,$, with Nambu-Goldstone bosons\, $\pib_j \equiv \pi^a_jT^a$  ($j=1,\cdots, N+1$), and covariant derivatives
$~
\dis
  D^\mu U_j = \partial^\mu U_j - i g_{j-1}^{~}\Ab_{j-1}^\mu U_j
                               + i g_j^{~}U_j \Ab_{j}^\mu \,. ~$
We further define the diagonal matrices of f-constants and couplings
\,$2 \CF_{N+1}
   \equiv {\rm diag}\(f_1^{~},f_2^{~},\cdots,f_{N+1}^{~}\)$\, and
\,$\CG_{N+1}
   \equiv {\rm diag}\(g_0^{~},g_1^{~},\cdots,g_N^{~}\)$\,,\,
and also the matrix $\CD_W$ whose entries $(\CD_W)_{ij}$ are $1$ (if $i=j$), $-1$ (if $i=j-1$), and zero otherwise.  These combine to form   
\beq
\QQB=\CF_{N+1}\CD_W\CG_{N+1}\, .
\label{eq:Q}
\eeq
The gauge boson mass-matrix $\MMW$ and
its dual $\MMWT$ can then be written as\,\cite{Sfetsos:2001qb}
\beq
\ba{ll}
\MMW = \QQB^T\QQB \,,~~ &
\MMWT = \QQB\,\QQB^T\,.
\ea
\label{eq:mwmwdq}
\eeq
Note that $\MMW$ and $\MMWT$ are real symmetric matrices,
and that $\QQB$ can be diagonalized by the
bi-orthogonal rotation,
\beq
\RRBT^T\QQB\RRB = \QQB^{\rm diag}  \equiv \MMB_W^{\rm diag}\, ,
\label{eq:qrrtdiag}
\eeq
which leads to
\beq
\RRB^T\MMW\RRB = (\QQB^{\rm diag})^T(\QQB^{\rm diag}) 
= (\MMB_W^2)^{\rm diag}~,
\label{eq:five}
\eeq
and
\beq
\RRBT^T\MMWT\RRBT = (\QQB^{\rm diag})(\QQB^{\rm diag})^T \equiv (\MMBT_W^2)^{\rm diag}\,~.
\eeq  
Since $ (\QQB_W^{\rm diag})^T(\QQB_W^{\rm diag})=(\QQB_W^{\rm diag})(\QQB_W^{\rm diag})^T$\,, 
we conclude that $\,(\MMW )^{\rm diag} =(\MMWT)^{\rm diag}$
and  the dual mass matrix \,$\MMWT$ has 
the same eigenvalues as \,$\MMW$\,.\,

Expanding the Lagrangian (\ref{eq:L}) yields
the gauge-Nambu-Goldstone boson mixing term,
\beq
\label{eq:A-GB}
{\cal L}_{\rm GB}^{\rm mix} ~=~
-{{\mathbf A}^{a}_{\mu}}^T\QQB^{T}
    \partial^\mu\Pi^a
~=~
-{\WWB^{a}_{\mu}}^T\MMB_{W}^{\rm \,diag}
    \partial^\mu\Pit^a  \,,
\eeq
with
\beq
\ba{ll}
\AAB^{a\mu}\,=\,
(A_0^{a\mu},\,A_1^{a\mu},\,\cdots,\,A_N^{a\mu}\,)^T,
~~&~~
\WWB^{a\mu}\,=\,
(W_0^{a\mu},\,W^{a\mu}_1,\,\cdots,\,W^{a\mu}_N\,)^T,
\\[2mm]
\Pi^{a}\,=\,
(\pi_1^{a},\,\pi^{a}_2,\,\cdots,\,\pi^{a}_{N+1}\,)^T,
~~&~~
\Pit^{a}\,=\,
(\pit_0^{a},\,\pit^{a}_1,\,\cdots,\,\pit^{a}_N\,)^T,
\ea
\eeq
where $\,\{\WWB^{a\mu}\}\,$
are mass-eigenbasis fields and
$\,\{\Pit^a\}\,$ are ``eaten'' Nambu-Goldstone fields,
which are connected to the site gauge bosons
$\,\{\AAB^{a\mu}\}\,$ and link Nambu-Goldstone bosons $\,\{\Pi^a\}\,$ via
\beq
\label{eq:A-PI-rotation}
\WWB^{a\mu} \,=\, \RRB^T\,\AAB^{a\mu} \,, ~~~~~~
\Pit^a   \,=\, \RRBT^T\,\Pi^a \,.
\eeq
Hence, the ``eaten'' Nambu-Goldstone modes are exactly aligned with the
``gauge boson'' mass-eigenstates of the dual moose.
The gauge-Nambu-Goldstone mixing (\ref{eq:A-GB}) can be
removed by the familiar
$R_\xi$ gauge-fixing term, 
\beq
\label{eq:GF}
\ba{l}
{\cal L}_{\rm gf} \,=\, \dis
\sum_{n=0}^N
- \f{1}{\,2\xi\,}\(F_n^a\)^2  \,,~~~~~~
\dis
F_n^a \,=\,
\dif_\mu W_n^{a\mu}  +
\xi M_{n}\pit_n^a  \,.
\ea
\eeq
So, relative to the mass $M_n$ of a mass-eigenstate gauge boon $W^{a\mu}_n$, the mass of the related ``eaten''  Nambu-Goldstone boson $\pit_n^a$ is
given by  $~M_{\pit^a_n}^2=\xi M_{n}^2\,$.


We may derive the gauge-sector quartic and cubic
vertices and express them in the mass-eigenbasis by expanding the appropriate terms of (\ref{eq:L}) 
\beqa
\label{eq:Lint-gauge}
{\cal L}_G^{\rm int} &=&
\sum_{j=0}^N\[
-\f{g_j}{2}\ep^{abc}\ov{F}_j^{a\mu\nu}A_{j\mu}^bA_{j\nu}^c
-\f{g_j^2}{4}\ep^{abc}\ep^{ade}A_j^{b\mu}A_j^{c\nu}A_{j\mu}^dA_{j\nu}^e
\]
\nonumber
\\[2mm]
&=&
-\f{\,G_3^{kmn}\,}{2}
 \ep^{abc}\ov{W}_k^{a\mu\nu}W^b_{m\mu}W^c_{n\nu}
-\f{\,G_4^{k\ell mn}\,}{4} \epsilon^{abc} \epsilon^{ade}
 W_k^{b\mu}W_{\ell}^{c\nu}W_{m\mu}^dW_{n\nu}^e \,,
\eeqa
using the notation 
$\,\ov{F}^{a\mu\nu}_j\equiv \dif^\mu A_j^{a\nu}-\dif^\nu A_j^{a\mu}\,$
and
$\,\ov{W}^{a\mu\nu}_j\equiv \dif^\mu W_j^{a\nu}-\dif^\nu W_j^{a\mu}\,$.\,
The quartic and cubic gauge couplings
$G_4^{k\ell mn}$ and $G_3^{kmn}$
are  defined as\,\cite{He:2004zr},
\begin{eqnarray}
  G_3^{nmk} &\equiv& \sum_j g_j \bbR_{j,n} \bbR_{j,m} \bbR_{j,k},
\label{eq:def3},
\\
  G_4^{nm\ell k} &\equiv& \sum_j g_j^2 
    \bbR_{j,n} \bbR_{j,m} \bbR_{j,\ell}\bbR_{j,k} .
\label{eq:def4}
\end{eqnarray}
Each of these multi-gauge-boson couplings is symmetric under exchange of any pair of indices and they are independent of the form of $\CD_W
$.

We may likewise derive from (\ref{eq:L}), the pure Nambu-Goldstone boson interactions,
\beqa
\label{eq:Lint-pi}
{\cal L}_\pi^{\rm int} &=&
\f{\Gt_4^{k\ell mn}}{6v^2}\[
\(\pit_k^a\dif_\mu\pit^a_{\ell}\)(\pit_m^b\dif^\mu\pit^b_n)-
\(\pit_k^a\pit^a_{\ell}\)(\dif_\mu\pit_m^b\dif^\mu\pit^b_n)+
{\cal O}(\pit_j^6)
\]
\eeqa
where
\beq
 \tilde{G}_4^{nm\ell k} 
  \equiv \sum_j \dfrac{v^2}{f_j^2} 
           \tilde{\bbR}_{j,n} \tilde{\bbR}_{j,m}
           \tilde{\bbR}_{j,\ell} \tilde{\bbR}_{j,k},
\label{eq:deft4}
\eeq
with factors of the weak scale  $v =\(\sqrt{2}G_F\)^{-1/2}$ inserted to make the couplings dimensionless.
Again, this coupling is Bose-symmetric under exchange of any pair of indices and its form does not depend on $\CD_W
$.  Note that $\tilde{G}_4$ vanishes in the 5D continuum limit ($N\to \infinity$) and that $\Gt_4^{nnmm}=1$ in the special case $\,N+1=1\,$, corresponding to the Higgsless standard model.

Finally, there are the interactions between Nambu-Goldstone modes and gauge bosons
 \beqa
 \label{eq:Lint-piW}
  {\cal L}_{\pi W}^{\rm int} &\!\!=\!\!&
 \f{1}{2}\Gt_3^{mnk}\,\ep^{abc}\pit^a_m\dif_\mu\pit^b_n W_k^{c\mu} + \ 
 (M_n +  M_m)\tilde{G}_{31}^{nmk}\, \ep^{abc} W^a_{\mu n} W^{b \mu}_{m} \pi^c_k \nonumber\\
 &+\!\!&\f{\,\Gt_{41}^{nm\ell k}\,}{3v}\!\[
 \pit_n^b\dif_\mu\pit_m^b\pit_{\ell}^a -
 \pit_m^b\pit_{\ell}^b\dif_\mu\pit_n^a
 \]\!W_k^{a\mu} 
  -\f{1}{4}
  [2\, \delta^{ab}\delta^{cd}-\delta^{ac}\delta^{bd}-\delta^{ad}\delta^{bc}]\,
  \Gt_{42}^{nm\ell k}\pit_n^a\pit_m^b W_k^{\mu c}W_{\mu \ell}^d + \cdots \,,
 \eeqa
The form of the interactions is general, but the specific definition of each coupling in terms of the $\RRB$ and $\RRBT$ matrices depends on $\CD_W
$ -- that is, on the structure of the model.  For the linear deconstructed model we are using as a benchmark
the quartic $\pi\pi\pi V$ coupling, and
$\pi\pi VV$ coupling \footnote{Note that we symmetrized $\ell \leftrightarrow k$ (anti-symmetrized $n
 \leftrightarrow m$ ) in the definition of $\tilde{G}_{42}^{nm\ell k}$ ($\tilde{G}_{31}^{nmk}$).} are defined as
\begin{eqnarray}
   \tilde{G}_{41}^{nm\ell k} 
  &\equiv& \sum_j \dfrac{v}{f_j} 
           \tilde{\bbR}_{j,n} \tilde{\bbR}_{j,m}
           \tilde{\bbR}_{j,\ell} 
           (g_j \bbR_{j,k} - g_{j-1} \bbR_{j-1,k}),
\label{eq:deft41}
  \\
  \tilde{G}_{42}^{nm\ell k} 
  &\equiv& -\frac{1}{2} \sum_j g_{j-1} g_j 
           \tilde{\bbR}_{j,n} \tilde{\bbR}_{j,m}
           (\bbR_{j-1,\ell} \bbR_{j,k} + \bbR_{j,\ell} \bbR_{j-1,k}),
\label{eq:deft42}
\end{eqnarray}
Similarly, the $VV\pi$ and $\pi\pi V$ couplings are defined in such models as
\begin{eqnarray}
  (M_n + M_m) \tilde{G}_{31}^{nmk}
    &\equiv& \frac{1}{2} \sum_j g_j g_{j-1} f_j 
    (\bbR_{j,n} \bbR_{j-1,m} - \bbR_{j-1,n} \bbR_{j,m})
   \tilde{\bbR}_{j,k}, 
\label{eq:deft31}
  \\
  \tilde{G}_3^{nmk} 
  &\equiv& \sum_j \tilde{\bbR}_{j,n} \tilde{\bbR}_{j,m}
           (g_j \bbR_{j,k} + g_{j-1} \bbR_{j-1,k}) .
\label{eq:deft3}
\end{eqnarray}
As we will discuss in Section \ref{sec:concl}, the results derived specifically for linear moose models in Section V can be generalized to apply to broader classes of models if one allows for a more general form of $\CD_W$.  

\section{General Longitudinal and Nambu-Goldstone Scattering Amplitudes
}

In this section, we derive the leading contributions (in powers of scattering energy) to the amplitudes for the $2 \to 2$ scattering of  gauge bosons and Nambu-Goldstone modes.   We specifically focus on the inelastic scattering where two bosons of KK level $n$ scatter into two modes of KK level $m$, where $n \neq m$.  Amplitudes involving only gauge boons are calculated in unitary gauge, while those involving one or more Nambu-Goldstone bosons are calculated in 't-Hooft-Feynman gauge; issues related to gauge invariance are addressed in Section V. We will apply the Equivalence Theorem to these results in Section \ref{ETsumsec} to derive sum rules constraining the bosons' 3-point and 4-point couplings. 

\subsection{Notation and Kinematics for $nn\to mm$ Scattering}

 We start by defining the kinematic variables for $nn\to mm$ KK-gauge-boson or Nambu-Goldstone-boson scattering.  In the center-of-mass frame, the 4-momenta of the incoming (1,2) and outgoing (3,4) states may be written as 
 \beqa
 p_1^{}=(\frac12 E,0,0,p),~~~   p_2^{}=(\frac12 E,0,0,-p),~~~
 p_3^{}=(\frac12 E,q\sin\theta,0,q\cos\theta),~~~ p_4^{}=(\frac12 E,-q\sin\theta,0,-q\cos\theta),
 \eeqa
 where
 \,$\frac12 E =\sqrt{p^2+M_n^2}=\sqrt{q^2+M_m^2}
        \equiv \f{1}{2}\sqrt{s}$\, and $\theta$ is the scattering angle.
 The corresponding polarization vectors for the incoming/outgoing states that are longitudinal gauge bosons are of the form:
 \beq
 \eph_{L1}^{}=\f{(p,0,0,\frac12 E)}{M_n},     \quad
 \eph_{L2}^{}=\f{(p,0,0,-\frac12 E)}{M_n},   \quad
 \eph_{L3}^{}=\f{(q,\frac12 E\sin\theta,0,\frac12 E\cos\theta)}{M_m},   \quad
 \eph_{L4}^{}=\f{(q,-\frac12 E\sin\theta,0,-\frac12 E\cos\theta)}{M_m}\,.
 \eeq
 For the states that are transverse gauge bosons, two possible sets of polarization vectors are
 \beq
 \ba{llll}
 \dis
 \eph_{T1}^{}=(0,1,0,0),        &\dis
 \eph_{T2}^{}=(0,1,0,0),        &\dis
 \eph_{T3}^{}=(0,\cos\theta,0,-\sin\theta),       &\dis
 \eph_{T4}^{}=(0,\cos\theta,0,-\sin\theta),
 \\[3mm]
 \eph_{T1'}^{}=(0,0,1,0),        &\dis
 \eph_{T2'}^{}=(0,0,1,0),        &\dis
 \eph_{T3'}^{}=(0,0,1,0),       &\dis
 \eph_{T4'}^{}=(0,0,1,0),
 \label{eq:secondpolar}
 \ea
 \eeq 
The Mandelstaam variables are defined as
 \beqa
 s&=& (p_1^{}+p_2^{})^2=  E^2 \,,
\nonumber\\
 t&=& (p_1^{}-p_3^{})^2= -(p^2+q^2)+2pq \cos\theta \,,
 \\
 u&=& (p_1^{}-p_4^{})^2= -(p^2+q^2)-2pq \cos\theta \,,
 \nonumber \eeqa
 and thus $\,s+t+u=2(M_n^2+M_m^2)$\,.  
 
 In addition, we will use the compact definitions $c \equiv \cos\theta$ and 
 \beqa
M_{\pm}^{2} &\equiv& M_m^{2}\pm M_n^{2}\,, \qquad\qquad M^4_\pm \equiv \left( M_\pm^2\right)^2\,,
\\
\Eb&\equiv&  E / M_n\,, \qquad\qquad\qquad {\Mb_j} \equiv M_j / M_n\,, \qquad\qquad\qquad  \Mb_\pm^2 \equiv  M_{\pm}^2 / M_n^2 \,,
\label{eq:bardeff}\\
\Eh&\equiv&  E / M_m\,, \qquad\qquad\qquad {\Mh_j} \equiv M_j / M_m\,,  \qquad\qquad\qquad  \Mh_\pm^2 \equiv  M_{\pm}^2 / M_m^2 \,.
\label{eq:hatdeff}
 \eeqa
to simplify the expressions for the scattering amplitudes.  Note that the ``bar'' signifies normalizing by the mass of the $n$th gauge boson KK mode (the incoming state in $nn \to mm$ scattering), while a ``hat'' signifies normalizing by the mass of the outgoing state of KK index $m$.

 The $nn\to mm$ scattering amplitude among gauge bosons (longitudinal or transverse)
 contains contributions arising from four classes of Feynman diagrams: four-point contact,
  $s$-channel,  $t$-channel, and 
 $u$-channel: 
\beq
 \TT_W [nn,mm] = \TT_{c} + \TT_{s} + \TT_{t} + \TT_{u}\,.
 \eeq
 The amplitude may be decomposed as follows
  \beq
 \TT_W [nn,mm]  = \ep^{abe}\ep^{cde}T_1 +
     \ep^{ace}\ep^{bde}T_2 +
     \ep^{ade}\ep^{bce}T_3
     \label{eq:Tnnmm}
 \eeq
 where the first (second, third) term receives contributions only from the contact and $s$ ($t$,$u$)-channel diagrams:
 \beq
 T_1 = T_{c1} + T_{s} \qquad
 T_2 = T_{c2} + T_{t}\,,\qquad
 T_3 =  T_{c3} + T_{u} \,.
 \eeq
Using the Jacobi identity
$~
\ep^{abe}\ep^{cde} + \ep^{ace}\ep^{dbe} + \ep^{ade}\ep^{bce} = 0
~$, we can rewrite (\ref{eq:Tnnmm}) as
\beq
\label{eq:Tnnmm-2}
\T_W[nn,mm] = 
    \ep^{ace}\ep^{bde} ( T_1 + T_2) +
    \ep^{ade}\ep^{bce} (-T_1 + T_3)\,.
\eeq
Furthermore, we will see that because $\,T_1\propto c\,,$ while $T_3 =T_2[c \to -c]$,
the combination $\,-T_1+T_3\,$ can be obtained from $\,T_1+T_2\,$
via a simple exchange $\,c\to -c\,$.\, Hence, to obtain the entire
amplitude $T_W[nn,mm]$ we only need to
compute $\,T_1+T_2$ explicitly.

The $nn\to mm$ scattering amplitudes involving Nambu-Goldstone bosons (and sometimes also transverse gauge bosons) may similarly be decomposed into contributions from four-point contact, $s$-channel,  $t$-channel, and  $u$-channel diagrams:
 \beq
\label{eq:Tpi-nnmm}
\ba{lcl}
\Tt_{\pi}[nn,mm] &=& \Tt_c + \Tt_s + \Tt_t + \Tt_u
\\[4mm]
&=& \ep^{abe}\ep^{cde} \Tt_1 +
    \ep^{ace}\ep^{bde} \Tt_2 +
    \ep^{ade}\ep^{bce} \Tt_3
    \ea
 \eeq
 Once again, the Jacobi identity can be used to simplify the result
 \beq
\Tt_{\pi}[nn,mm] = 
    \ep^{ace}\ep^{bde} (\Tt_1+\Tt_2) +
    \ep^{ade}\ep^{bce} (-\Tt_1+\Tt_3)\,.
\eeq
 and we will see that $-\tilde{T}_1 + \tilde{T}_3$ can be obtained from $\tilde{T}_1 + \tilde{T}_2$ by the exchange $\,c\to -c$ so that only $\tilde{T}_1 + \tilde{T}_2$ need be calculated explicitly.  However, in this case it is worth noting that there is an alternative way to express the amplitude
 \beq
 \Tt_{\pi}[nn,mm] =  \delta^{ab}\delta^{cd} \Tbt_{1}
          + \delta^{ac}\delta^{bd} \Tbt_{2}
          + \delta^{ad}\delta^{bc} \Tbt_{3} 
 \eeq
 where
 \beq
 \widetilde{\overline{T}}_1 = \widetilde{T}_2 + \widetilde{T}_3
\,, ~~~~
\widetilde{\overline{T}}_2 = \widetilde{T}_1 - \widetilde{T}_3\,, ~~~~
\widetilde{\overline{T}}_3 = -\widetilde{T}_1 - \widetilde{T}_2\,.
 \eeq
The four-point contact interactions are more easily calculated in terms of the $\Tbt$ while the $s$-,$t$-, and $u$-channel diagrams are more naturally written in terms of the  $\Tt$ since only the $s$-channel diagram contributes to $\Tt_1$ (and so on). Hence, it is convenient to perform the calculation of  $\tilde{T}_1 + \tilde{T}_2$ by writing
 \beq
 \Tt_1+\Tt_2 = \Tt_s + \Tt_t - \Tbt_{c3}
 \eeq
where $\widetilde{\overline{T}}_{c3}$ denotes the contact contribution to the
amplitude $\widetilde{\overline{T}}_3$.

We will now calculate the scattering amplitudes for all the relevant processes involving gauge and Nambu-Goldstone bosons and  extract the pieces from which sum rules may be derived via application of the Equivalence Theorem.
          
\subsection{Amplitudes with no transverse gauge bosons}

\subsubsection{
Purely Longitudinal Gauge Amplitude $W_{nL}^aW_{nL}^b\to W_{mL}^cW_{mL}^d$ ($n,m\geqq 0$)
}

We begin by considering the longitudinal gauge boson scattering process
$W_{nL}^aW_{nL}^b\to W_{mL}^cW_{mL}^d$ ($n,m\geqq 0$),
where $n=m$ corresponds to elastic scattering and $n\neq m$
corresponds to inelastic scattering.  We will display the results of this calculation in detail 
 to show how the ideas of the last section play out in a specific case; the results for other
processes will also be summarized.
The scattering amplitude contains contributions from the 4-boson contact diagram and
the diagrams in the $(s,t,u)$ channels,
\beqa
\T_W[nn,mm] &=& \T_c + \T_s + \T_t + \T_u
\nonumber
\\[4mm]
&=& \ep^{abe}\ep^{cde} T_1 +
    \ep^{ace}\ep^{bde} T_2 +
    \ep^{ade}\ep^{bce} T_3\,,
   \nonumber \\ [4mm]
&=&
    \ep^{ace}\ep^{bde} ( T_1 + T_2) +
    \ep^{ade}\ep^{bce} (-T_1 + T_3)\,.
\eeqa

We compute the expressions for the full scattering amplitudes $T_j$ to be
\beqa
 T_1 &=& {T_{c1}} + {T_s}
 \nonumber \\
 &=& {\f{\,c\,pq\,}{M_n^2M_m^2} \[-G_4^{nnmm}E^2\]} +  {\f{\,c\,pq\,}{M_n^2M_m^2}\[ \f{1}{4}\(3E^2\!-\!4q^2\)\(3E^2\!-\!4p^2\) \sum_k\f{\,G_3^{nnk}G_3^{mmk}\,}{\,E^2\!-\!M_k^2\,}  \]  }\!,
 \eeqa
 \beqa
 T_2 &=& {T_{c2}} + {T_t}
 \nonumber \\[3mm]
 &=& {\f{\,G_4^{nnmm}}{\,M_n^2M_m^2\,}\[\f{1\!-\!c^2}{16}E^4 - \f{1}{4}tE^2\]}
 \nonumber \\[3mm]
 && + {\sum_k \f{(G_3^{nmk})^2}{(-t\!+\!M_k^2)M_n^2M_m^2}\[
 \f{3\!+\!c^2}{32}E^6 - \f{(5\!-\!c^2)c}{8}pqE^4 -\f{3(1\!+\!3c^2)}{16}M_+^2E^4 +\(2M_+^2\!-\!\f{M_-^4}{2M_k^2}\)cpqE^2 \right.}
 \nonumber \\[3mm]
 && \dis \hspace*{10mm}
 +\f{1\!+\!c^2}{16M_k^2}M_-^4E^4 +\(3c^2M_n^2M_m^2 +\f{1\!+\!4c^2}{4}(M_n^4\!+\!M_m^4)-\f{M_-^4M_+^2}{4M_k^2}\)E^2 
 \\[3mm]
 && \dis {\left. \hspace*{10mm}
 +2cM_n^2M_m^2pq  +  \f{M_n^2M_m^2M_-^4}{M_k^2} - M_n^2M_m^2M_+^2 \] }\!, \nonumber
  \\[3mm]
  T_3 &=& T_{c3}+T_u ~=~ T_2[t\to u,c\to -c] \,.
 \eeqa
Expanding these amplitudes $T_j$ in the high-energy limit, we may identify 
all the contributions proportional to $E^4$, $E^2$ and $E^0$,
\beqa
\label{eq:T1nnmm}
T_1 &\!\!\!=\!\!\!&
\dis \f{c}{4M_n^2M_m^2}\left\{\! G_4^{nnmm}\!
\(E^4 \!-\! 2M_+^2E^2 - 2M_-^4E^0\) \right.
\\
&&
\left.\hspace*{17mm}
-\!\sum_k\! G_3^{nnk}G_3^{mmk}
\[ E^4 \!+\! M_k^2E^2 +\(M_k^4-6(M_n^4 \!+\! M_m^4)\)E^0\]\right\}\!
+O\!\(\f{1}{E}\)\,,
\nonumber
\\[2mm]
T_2  &\!\!\!=\!\!\!&
\dis \f{1}{4M_n^2M_m^2}\left\{G_4^{nnmm}\!
\[-\f{(1-c)(3+c)}{4}E^4+(1-c)M_+^2E^2 - cM_-^4E^0 \] \right.
\nonumber
\\[3mm]
&\!\!\!\!\!\!& \left. \!\!\!
+\sum_k \(\!G_3^{nmk}\!\)^2\!
\[\f{(1\!-\!c)(3\!+\!c)}{4}E^4 +
  \(\!\! 4cM_+^2 \!+\! \f{1\!-\!c}{2}\f{M_-^4}{M_k^2}
      \!-\! \f{3\!+\!c}{2}M_k^2\!\)\! E^2 +
\nonumber  \right.\right.
\\[2mm]
&&
\hspace*{25mm}
\(
\f{3\!+\!c}{1\!-\!c}M_k^4+\f{c^2\!-\!6c\!-\!3}{1\!-\!c}M_+^2M_k^2-
(1\!+\!c)\f{M_-^4M_+^2}{M_k^2}\,+
\right.
\label{eq:T2nnmm}
\\[2mm]
&&
\hspace*{25mm}
\left.\left.\left.
+\f{(1\!+\!c)^2}{1\!-\!c}M_+^4 - 4cM_n^2M_m^2
\,\)\! E^0\,
\]~\right\} +O\!\(\f{1}{E}\)\,,
\nonumber
\\[3mm]
T_3  &\!\!\!=\!\!\!& T_2[c \to -c]\ .
\label{eq:T3nnmm}
\eeqa

As expected, $\,T_1\propto c\,$ while $T_2$ and $T_3$ are related as in (\ref{eq:T3nnmm}).
 Hence, to obtain the entire amplitude $T_W[nn,mm]$ we only need to
compute $\,T_1+T_2$. 
From (\ref{eq:T1nnmm})-(\ref{eq:T3nnmm}), we now display the separate terms arising in 
$\,T_1+T_2\,$ at orders $E^4$, $E^2$ and $E^0$, respectively.
\beqa
\(T_1\!+\!T_2\)_{E^4}
&\!\!\!=\!\!\!&
\f{E^4}{4M_n^2M_m^2}
\left\{\[G_4^{nnmm}- \sum_k\! G_3^{nnk}G_3^{mmk}\]c ~+ \right.
\nonumber
\\[2mm]
&& \left.\hspace*{1.9cm}
\[G_4^{nnmm}- \sum_k\! \(G_3^{nmk}\)^2\]\f{-(1\!-\!c)(3\!+\!c)}{4}
\right\},
\label{eq:WL-E4}
\\[4mm]
\(T_1\!+\!T_2\)_{E^2}
&\!\!\!=\!\!\!&
\f{E^2}{4M_n^2M_m^2}
\[ (1\!-\!3c)M_+^2G_4^{nnmm}
-\sum_k\! G_3^{nnk}G_3^{mmk}M_k^2c ~+ \right.
\nonumber
\\[2mm]
&&  \left. \hspace*{1.7cm}
+\sum_k \(G_3^{nmk}\)^2
\( 4cM_+^2+\f{1\!-\!c}{2}\f{M_-^4}{M_k^2} - \f{3\!+\!c}{2}M_k^2 \)
\],
\label{eq:WL-E2}
\\[4mm]
\(T_1\!+\!T_2\)_{E^0}
 &\!\!\!=\!\!\!&
\f{E^0}{4M_n^2M_m^2} \!\!
\left\{
\!G_4^{nnmm}\!\(\!-3cM_-^4\) \!+\!
\!\sum_k\! G_3^{nnk}G_3^{mmk}\!\!\[-cM_k^4\!+\!6c(M_n^4\!+\!M_m^4)\] +
\right.
\nonumber
\\[2mm]
&&  \hspace*{1.7cm}
\sum_k\!\!\(\!G_3^{nmk}\!\)^2\!\!\[
\f{3\!+\!c}{1\!-\!c}M_k^4\!+\!\f{c^2\!-\!6c\!-\!3}{1\!-\!c}M_+^2M_k^2
\!-\!(1\!+\!c)\f{M_+^2M_-^4}{M_k^2}\,+\right.
\label{eq:WL-E0}
\\[2mm]
&& \hspace*{1.7cm} \left.\left.
\f{(1+c)^2}{1-c}M_+^4 -4cM_n^2M_m^2
\]\right\} .
\nonumber
\eeqa
We will apply the Equivalence Theorem to these expressions shortly, in order to derive sum rules.  First, however, we need to compute the Nambu-Goldstone scattering amplitude to which the Equivalence Theorem will relate our present results.

\subsubsection{
Purely Nambu-Goldstone Amplitude
$\pit_{n}^a\pit_{n}^b\to \pit_{m}^c\pit_{m}^d$ ($n,m\geqq 0$)
}

Now we consider the related Nambu-Goldstone boson scattering process
$\pit_{n}^a\pit_{n}^b\to \pit_{m}^c\pit_{m}^d$ ($n,m\geqq 0$).
Beginning from the Lagrangian (\ref{eq:Lint-pi}) we directly compute the Nambu-Goldstone boson  contact
contributions,
\beq
\label{eq:Tpi-contact}
\dis
\Tbt_{c1} = \f{\Gt_4^{nnmm}}{v^2}\[s -\f{2}{3}M_+^2\]  ,~~~~
\Tbt_{c2} = \f{\Gt_4^{nnmm}}{v^2}\[t -\f{2}{3}M_+^2\]    ,~~~~
\Tbt_{c3} = \f{\Gt_4^{nnmm}}{v^2}\[u -\f{2}{3}M_+^2\]    ,
\eeq
with $\,s+t+u = 2M_+^2\,$.
In (\ref{eq:Tpi-contact}) the quartic Nambu-Goldstone boson coupling
$\Gt_4^{nnmm}$ is given by the expression in Eq. (\ref{eq:deft4}), 
which vanishes in the 5D continuum limit ($N\to \infinity$).

Starting from the Lagrangian, we can also calculate the contributions from $W_k$ exchange in the
$(s,t,u)$-channels,
\beqa
\hspace*{-5mm}
\Tt_{stu}[nn,mm] &\!\!=\!\!&
\ep^{ace}\ep^{bde}
\sum_k\f{1}{4}\(
\Gt_3^{nnk}\Gt_3^{mmk}\f{u\!-\!t}{s\!-\!M_k^2} +
\(\Gt_3^{nmk}\)^2 \f{s\!-\!u}{-t\!+\!M_k^2}
\) +
\nonumber\\
&&
\ep^{ade}\ep^{bce}
\sum_k\f{1}{4}\(
\Gt_3^{nnk}\Gt_3^{mmk}\f{-u\!+\!t}{s\!-\!M_k^2} +
\(\Gt_3^{nmk}\)^2 \f{s\!-\!t}{-u\!+\!M_k^2}
\)  ,
\label{eq:T-pi-stu}
\eeqa
where we have adopted `t Hooft-Feynman gauge ($\xi=1$); extension 
to arbitrary $\xi$ will be discussed
in section V.
Recalling that taking $c \to -c$ corresponds to exchanging $t$ with $u$, we see that  the two terms in
$~\ep^{ace}\ep^{bde}\[\cdots\]~$ and
$~\ep^{ade}\ep^{bce}\[\cdots\]~$ are interchanged under $~c\to -c~$, as expected.

We can combine the preceeding to obtain 
 the total Nambu-Goldstone boson amplitude as below
\beq
\label{eq:Tpi-nnmmSum}
\ba{l}
\Tt_\pi [nn,mm]
\\[4mm]
~=~ \dis
\ep^{ace}\ep^{bde}\[
\f{\Gt_4^{nnmm}}{v^2}\(\! -u \!+\!\f{2}{3}M_+^2 \)+
\sum_k\f{1}{4}\(
\Gt_3^{nnk}\Gt_3^{mmk}\f{u\!-\!t}{s\!-\!M_k^2} +
\(\Gt_3^{nmk}\)^2 \f{s\!-\!u}{-t\!+\!M_k^2}
\)\]
+
\\[7mm]
\dis ~~~~~~
\ep^{ade}\ep^{bce}\[
\f{\Gt_4^{nnmm}}{v^2}\(\! -t \!+\!\f{2}{3}M_+^2 \)+
\sum_k\f{1}{4}\(
\Gt_3^{nnk}\Gt_3^{mmk}\f{-u\!+\!t}{s\!-\!M_k^2} +
\(\Gt_3^{nmk}\)^2 \f{s\!-\!t}{-u\!+\!M_k^2}
\)\] .
\ea
\eeq
Note that, unlike the longitudinal gauge boson amplitude $T_W[nn,mm]$,
the Nambu-Goldstone boson amplitude $\Tt_\pi [nn,mm]$ contains no $E^4$-term
and all nonzero terms grow with ${\cal O}(E^2)$ at most.
Furthermore, in the 5D continuum limit, terms growing like $E^2$ in 
the Nambu-Goldstone boson amplitude  $\Tt_\pi [nn,mm]$
vanish and the leading terms in the amplitude go as $E^0$.

Finally, we expand (\ref{eq:Tpi-nnmmSum}) for large-$E$ and derive the
asymptotic amplitudes at ${\cal O}(E^2)$ and ${\cal O}(E^0)$, respectively,
\vspace*{4mm}
\beq
\label{eq:Tpi-E2}
\ba{l}
\hspace*{-68.3mm}
\dis\(\Tt_1 + \Tt_2\)_{E^2} =
\f{\Gt_4^{nnmm}}{2v^2} (1 + c)\, E^2 \,,
\ea
\eeq
\vspace*{1.5mm}
\beq
\label{eq:Tpi-E0}
\ba{l}
\dis\(\Tt_1 + \Tt_2\)_{E^0} = \[
-\f{\Gt_4^{nnmm}M_+^2}{2v^2}\(\f{1}{3}\!+\!c\) +
\sum_k\f{1}{4}\(
\Gt_3^{nnk}\Gt_3^{mmk}(-c) + \(\Gt_3^{nmk}\)^2\f{3\!+\!c}{1\!-\!c}
\) \] \,.
\ea
\eeq
%

\subsection{Amplitudes with one transverse gauge boson}

\subsubsection{
Gauge Amplitude $W_{nT}^a W_{nL}^b\to W_{mL}^c W_{mL}^d$
($n,m\geqq 0$) }

 Next, we consider the scattering process
 $W_{nT}^a W_{nL}^b\to W_{mL}^c W_{mL}^d$\,, with one transversely polarized external field. 
 The terms in these amplitudes that grow with energy are of order $E^3$ and $E^1$. Expanding the amplitudes out to order $E^1$, we obtain
 \beqa
 \hspace*{-10mm}
 T_1[TL,LL] &\!\!\!=\!\!\!&
 \f{\sqrt{1\!-\!c^2}}{2\sqrt{2}M_nM_m^2}\!
 \left\{\!
  G_4^{nnmm}\!\(\!\f{E^3}{2} \!-\! M_+^2E\!\) \!-\!
  \sum_k \!G_3^{nnk}G_3^{mmk}\!\[E^3\!+\!(M_k^2\!-\!2M_n^2)E\]\!
 \right\}
 \label{eq:TLLL-T1}
 \\[3mm]
 \hspace*{-10mm}
 T_2[TL,LL] &\!\!\!=\!\!\!&
 \f{\sqrt{1\!-\!c^2}}{4\sqrt{2}M_nM_m^2}\!
 \left\{
 G_4^{nnmm}\(\f{1\!+\!c}{2}E^3-M_+^2E\) +
 \right.
 \label{eq:TLLL-T2}
 \\[1mm]
 \hspace*{-10mm} && \left.  \hspace*{24mm}
 \sum_k\(G_3^{nmk}\)^2\[
 \f{1\!-\!c}{2}E^3 - \(M_k^2-(5M_m^2+M_n^2)+\f{M_-^4}{M_k^2}\)E
 \] \right\}
 \nonumber
 \\[3mm]
 \hspace*{-10mm}
 T_3[TL,LL] &\!\!\!=\!\!\!& T_2[TL,LL](c\to -c) \,.
 \label{eq:TLLL-T3}
 \eeqa
 With these we derive the combination $T_1+T_2$ at ${\cal O}(E^3)$ and ${\cal O}(E^1)$,
 \beqa
 \label{eq:TLLL-T1+T2-E3}
 \hspace*{-4mm}
 (T_1\!+\!T_2)_{E^3} &\!\!\!=\!\!\!\!&
  \f{\sqrt{1\!-\!c^2}}{8\sqrt{2}M_nM_m^2}\!\[
  4\!\(\!G_4^{nnmm}\!-\!\sum_k\! G_3^{nnk}G_3^{mmk}\!\) \!-\!
  \(\!G_4^{nnmm}\!-\!\sum_k\! \(G_3^{nmk}\)^2\!\)\!\!(1\!-\!c)
  \]\!\! E^3
  \nonumber\\
 \hspace*{-4mm} &\!\!\!\!\!\!\!&
 \\[3mm]
 \label{eq:TLLL-T1+T2-E1}
 \hspace*{-4mm}
 (T_1\!+\!T_2)_{E^1} &\!\!\!=\!\!\!\!&
 \f{\sqrt{1\!-\!c^2}}{4\sqrt{2}M_nM_m^2}\!\[
  G_4^{nnmm}(-3M_+^2) +
  \sum_k\! G_3^{nnk}G_3^{mmk}\,2\(2M_n^2-M_k^2\)  \right.
  \nonumber\\
  \hspace*{-4mm} &\!\!\!\!\!\!\!&  \hspace*{25mm} \left.
  -\sum_k\(G_3^{nmk}\)^2\(M_k^2-(5M_m^2+M_n^2)+\f{M_-^4}{M_k^2}\)
 \] E^1\,.
 \eeqa
 %

\subsubsection{
Nambu-Goldstone Amplitude
$W_{nT}^a\pit_{n}^b\to\pit_{m}^c\pit_{m}^d$ ($n,m\geqq 0$) }

 For the related Nambu-Goldstone boson process
 $W_{nT}^a\pit_{n}^b\to\pit_{m}^c\pit_{m}^d$, no diagram grows as $E^3$, and the
 only diagram that contributes at ${\cal O}(E^1)$ is the contact diagram
 including the $W-\pit-\pit-\pit$ coupling from (\ref{eq:Lint-piW}); all other
 diagrams are of ${\cal O}(E^0)$ or smaller.
 Computing the amplitude at ${\cal O}(E^1)$, we find
 \beqa
 \Tt[W_T\pit,\pit\pit]_{E^1} &\!\!=\!\!&
  \f{\Gt_{41}^{mmnn}}{2\sqrt{2}v}
  \(\d^{ac}\d^{bd} - \d^{ad}\d^{bc}\)\sqrt{1\!-\!c^2}\, E \,,
 \eeqa
 which gives
 \beq
 \label{eq:Tpipipi-T1+T2-E1}
\( \Tt_1 + \Tt_2\)_{E} ~=~ \f{\Gt_{41}^{mmnn}}{2\sqrt{2}v}
                                \sqrt{1\!-\!c^2}\, E \,.
 \eeq
 For $\pit_{n}^a\pit_{n}^b\to W_{mT}^c\pit_{m}^d$, one obtains the same result with $n \leftrightarrow m$.

\subsection{Amplitudes with two transverse gauge bosons}

\subsubsection{
Gauge Amplitude $W_{nL}^a W_{nL}^b\to W_{mT}^c W_{mT}^d$
($n,m\geqq 0$) }

When we compute the amplitude for $W_{nL}^aW_{nL}^b\to W_{mT}^cW_{mT}^d$ employing 
the transverse polarization vectors $\eph_{T3}^{}$ and $\eph_{T4}^{}$ from (\ref{eq:secondpolar}), we find the respective contributions of the contact diagram, the $s$-channel diagram, and the $t$-channel diagram to be
 \beqa
 \TT_{c1} &=& 0 \,,
 \\
 \TT_{c2} &=& G_4^{nnmm}\[\f{1+c^2}{4}\Eb^2-1\] \,,
\\
 \TT_s &=& \ep^{abe}\ep^{cde}
 \sum_k G_3^{nnk}G_3^{mmk}
 \[-\f{1}{2}\Eb^2-\f{1}{2}\Mb_k^2 +\Mb_m^2\]c\,,
\\
 \TT_t &=& \ep^{ace}\ep^{bde}
 \sum_k (G_3^{nmk})^2\[
 -\f{(1\!-\!c)^2}{4}\Eb^2 + \f{1\!-\!c\,}{2}\Mb_k^2
 -\f{1\!+\!c\,}{2}\f{\Mb_-^4}{\Mb_k^2}
 \]
 \eeqa
 where the ``bar'' signifies normalizing by the mass of the $n$th gauge boson KK mode, as in Eq. (\ref{eq:bardeff}). Thus we deduce,
 \beqa
 \label{eq:LL-TT-E2}
 \hspace*{-4mm}
 (T_1\!+\!T_2)_{E^2}^{} &\!\!\!=\!\!\!&
 \[G_4^{nnmm}\!-\!\sum_k(G_3^{nmk})^2\]\!\f{1\!+\!c^2}{4}\Eb^2 +
 \sum_k\[(G_3^{nmk})^2\!-\!G_3^{nnk}G_3^{mmk}\]\f{c}{2}\Eb^2\,,
 \\[3mm]
 \hspace*{-4mm}
 (T_1\!+\!T_2)_{E^0}^{} 
 &\!\!\!=\!\!\!&
 -\[G_4^{nnnn}+\sum_k\hf(G_3^{nmk})^2
 \(-\Mb_k^2+\f{\Mb_-^4}{\Mb_k^2}\)\]
 \nonumber\\
 &\!\!\!\!\!\!&
 -\sum_k\[G_3^{nnk}G_3^{mmk}\(\hf\Mb_k^2-\Mb_m^2\)
          +(G_3^{nmk})^2\f{1}{2}\(\Mb_k^2+\f{\Mb_-^4}{\Mb_k^2}\)\]\!c
          \,.
 \label{eq:LL-TT-E0}
 \eeqa

On the other hand, if we perform the calculation employing the 
 transverse polarization vectors $\eph_{T3'}^{}$ and $\eph_{T4'}^{}$ instead, we obtain
  \beqa
 \label{eq:LL-TT'-E2}
 \hspace*{-8mm}
 (T_1'\!+\!T_2')_{E^2}^{} \!&\!\!\!=\!\!\!&
 \f{1}{2}\[G_4^{nnmm}\!-\!\sum_k(G_3^{nmk})^2\]\!\Eb^2 +
 \sum_k
\[(G_3^{nmk})^2\!-\!G_3^{nnk}G_3^{mmk}\]  \f{c}{2}\, \Eb^2\,,
 \\[3mm]
 \hspace*{-8mm}
 (T_1'\!+\!T_2')_{E^0}^{} 
 \!&\!\!\!=\!\!\!&\!
 -\[\!G_4^{nnmm}\!+\!\sum_k\!(G_3^{nmk})^2\!\(\Mb_-^2\!-\!\Mb_k^2\!\)\!\]
 \!\!-\!\sum_k\!\[\!G_3^{nnk}G_3^{mmk}\!\(\f{\Mb_k^2}{2}\!-\!\Mb_m^2\)
          \!+\! (G_3^{nmk})^2\Mb_-^2\!\]\!c\,.
 \nonumber\\
 \label{eq:LL-TT'-E0}
 \eeqa
We will see that the different results for the alternative polarization yield additional sum rules at ${\cal O}(E^0).$

Performing the related calculation for the inverse process $W_{nT}^aW_{nT}^b\to W_{mL}^cW_{mL}^d$
using the transverse polarization vectors $\eph_{T1}^{}$ and $\eph_{T2}^{}$ yields results of the same form as (\ref{eq:LL-TT-E2}) and (\ref{eq:LL-TT-E0}) but with $\Eb \to \Eh$ and $\Mb \to \Mh$.  That is, one still normalizes by the mass of the longitudinal external boson, but this is now an outgoing state of index $m$ as Eq. (\ref{eq:hatdeff}).   Likewise, if we use the alternative polarization vectors $\eph_{T1'}^{}$ and $\eph_{T2'}^{}$ we obtain a result of the same form as (\ref{eq:LL-TT'-E2}) and (\ref{eq:LL-TT'-E0}) but with $\Eb \to \Eh$ and $\Mb \to \Mh$.

\subsubsection{
Nambu-Goldstone Amplitude
$\pit_{n}^a\pit_{n}^b\to W_{mT}^cW_{mT}^d$ ($n,m\geqq 0$) }

When we compute the Nambu-Goldstone boson amplitude
 $\pi_{n}^a\pi_{n}^b\to W_{mT}^cW_{mT}^d$\,,
using the transverse polarization vectors
 $\eph_{T3}^{}$ and $\eph_{T4}^{}$,
we find that the relevant contact diagram contribution is of
 ${\cal O}(E^0)$:
 \beq
  \Tbt_{c3} = -\Gt_{42}^{nnmm} = \, - \Tbt_{c1} - \Tbt_{c2} \,.
 \eeq
The $s$-channel and $t$-channel contributions are also of ${\cal O}(E^0)$: 
 \beqa
 \Tt_{1s} &=& \sum_k \Gt_3^{nnk}G_3^{mmk}\f{\,-c\,}{2} \,,
 \\
 \Tt_{2t} &=& \sum_k (\Gt_3^{nkm})^2\f{1+c}{2} \,.
 \eeqa
Note that the $s$-channel contribution includes both pure-gauge and gauge-Nambu-Goldstone couplings, while the $t$-channel result depends only on gauge-Nambu-Goldstone couplings and involves a sum over the KK-Nambu-Goldstone mode index $k$ (rather than a KK-gauge boson index).   From the above equations, we find
 \beqa
 \label{eq:PiPi-TT}
 \Tt_1+\Tt_2 &=&
  \Gt_{42}^{nnmm} + \sum_k\Gt_3^{nnk}G_3^{mmk}\f{\,-c\,}{2}
   + \sum_k(\Gt_3^{nkm})^2\f{1\!+\!c}{2} + {\cal O}(E^{-2})\,.
 \eeqa

Repeating the calculation with the transverse polarization vectors
 $\eph_{T3'}^{}$ and $\eph_{T4'}^{}$ instead yields
 \beqa
 \label{eq:PiPi-TT'}
 \Tt_1'+\Tt_2' &=&
  \Gt_{42}^{nnmm} + \sum_k\Gt_3^{nnk}G_3^{mmk}\f{\,-c\,}{2}
    + {\cal O}(E^{-2})\,.
 \eeqa

 Performing the related calculation for the inverse process 
 $W_{nT}^aW_{nT}^b\to \pit_{m}^c\pit_{m}^d$ with polarization vectors $\eph_{T1}^{}$ and $\eph_{T2}^{}$  ($\eph_{T1'}^{}$ and $\eph_{T2'}^{}$) yields the same result as in Eq. (\ref{eq:PiPi-TT}) (Eq. (\ref{eq:PiPi-TT'})) but with $n \leftrightarrow m$.  

\subsection{Amplitudes with three transverse gauge bosons}
\label{sub:LTTT}

\subsubsection{
Gauge Amplitude $W_{nL}^a W_{nT}^b\to W_{mT}^c W_{mT}^d$
($n,m\geqq 0$) }

We first compute the amplitude for $W_{nL}^aW_{nT}^b\to W_{mT}^cW_{mT}^d$
using the transverse polarization vectors $(\eph_{T2},\,\eph_{T3}^{},\,\eph_{T4}^{})$ from (\ref{eq:secondpolar}).  We find the respective contributions of the contact diagram, the $s$-channel diagram, and the $t$-channel diagram to be
\beqa
  \TT_{c1} &=& 0 \,,
 \\
 \TT_{c2} &=& G_4^{nnmm}\[-\hf c\sqrt{1\!-\!c^2}\,\Eb\] \,,
 \\
 \TT_s &=& \ep^{abe}\ep^{cde}
 \sum_k G_3^{nnk}G_3^{mmk}\[\sqrt{1\!-\!c^2}\,\Eb\],
 \\
 \TT_t &=& \ep^{ace}\ep^{bde}
 \sum_k (G_3^{nmk})^2\[\(-1+\f{c}{2}\)\sqrt{1\!-\!c^2}\,\Eb\]\,.
  \eeqa
Combining these and keeping all terms of order $E^0$ or involving positive powers of $E$, yields
\beqa
 \label{eq:LT-TT}
 \hspace*{-6mm}
 T_1 + T_2  &\!\!\!=\!\!\!&
 \[\!G_4^{nnmm}\!-\!\!\sum_k(G_3^{nmk})^2\!\]\!\f{-c}{2}\sqrt{1\!-\!c^2}\,\Eb
 + \sum_k\!\[\!G_3^{nnk}G_3^{mmk}\!-\!(G_3^{nmk})^2\!\]\!\sqrt{1\!-\!c^2}\,\Eb
 \,,
 \eeqa
 where the ${\cal O}(E^0)$ contribution vanishes.  

We have also computed the amplitude for
 $W_{nL}^aW_{nT'}^b\to W_{mT'}^cW_{mT'}^d$ with the alternative 
 transverse polarizations, $(\eph_{T2'},\,\eph_{T3'}^{},\,\eph_{T4'}^{})$.
 Keeping terms of order $E^0$ or involving positive powers of $E$, we derive the final result,
 \beqa
 \label{eq:LT-T'T'}
 \hspace*{-6mm}
 T_1' + T_2'  &=&
 \sum_k\!\[\!G_3^{nnk}G_3^{mmk}\!-\!(G_3^{nmk})^2\!\]\!\sqrt{1\!-\!c^2}\,\Eb
 \,.
 \eeqa
 %

\subsubsection{
Nambu-Goldstone Amplitude
$~\pit^a W_{nT}^b\to W_{mT}^c W_{mT}^d$ ($n,m\geqq 0$) }

 Due to the absence of a 
 $\pit$-$W$-$W$-$W$ contact vertex, the leading term in the amplitude for $\pit W_{nT}\to W_{mT} W_{mT}$
 starts at ${\cal O}(1/E)$, i.e.,
 \beqa
 \Tt [\pit^a W_{nT}^b\to W_{mT}^c W_{mT}^d] &=& {\cal O}(1/E) \,.
 \eeqa
 We will find that this is of the same order as the $\f{M_{W_n}}{E}$ suppressed $B$-term 
  in the Equivalence Theorem (see Section \ref{subsec41}).

\section{General Sum Rules from the Equivalence Theorem}
\label{ETsumsec}

 In this section we use the Equivalence Theorem (ET) to
 derive general sum rules that relate the masses of the KK modes to the couplings among the gauge and Nambu-Goldstone bosons. These sum rules, which are collected in Table \ref{tab:sumrules}, are general in several senses.
  
 First, they apply to inelastic $nn \to mm$ scattering, not just to the more restricted case of elastic $nn\to nn$ scattering.  We show that some of the relations do have counterparts for elastic scattering, but others can be derived only by looking at inelastic processes.  
  
  Second, they arise directly in any consistent gauge theory in 4D or 5D, rather than being imposed from the outside like the 5D elastic sum rules in \cite{Csaki:2003dt}.  Moreover, while we have taken a linear deconstructed model as our benchmark, the form of the relations transcends this.  The sum rules relate  the various  $G$'s and $\tilde{G}$'s, defined simply as the coefficients of the multi-boson contact interactions arising from the Lagrangian kinetic energy terms in Eqs. (\ref{eq:Lint-gauge}), (\ref{eq:Lint-pi}) and (\ref{eq:Lint-piW}).  As mentioned in Section \ref{sec:higgsless}, in models other than a linear mooose, the detailed expansion of the pion-gauge-boson couplings in terms of the matrices $\RRB$ and $\RRBT$ will differ from what is written in Eqs. (\ref{eq:deft41}) - (\ref{eq:deft3}).  But that will not change the form of the sum rules as written in Table \ref{tab:sumrules}.  This is because all of our sum rules are derived from the ET which relies only on the spontaneously broken gauge symmetry of the Lagrangian.
  
Third, they apply directly to 4D deconstructed models and, when the appropriate limit is taken, to 5D continuum models as well.  The major difference between the continuum ($N \to \infty$) limit and the deconstructed theories with finite $N$ is that the four-pion coupling 
$\Gt_4$ and the $\pi\pi\pi W$ coupling $\Gt_{41}$ vanish in the continuum limit.  As a result, the order $E^2$ terms in the scattering amplitudes cancel completely only in 5D theories \,\cite{SekharChivukula:2001hz},
while for deconstructed theories order  $E^2$-terms originating from the
nonzero quartic Nambu-Goldstone boson coupling $\Gt_4^{nnmm}$ remain and
are suppressed by a factor $1/(N\!+\!1)$ \cite{Chivukula:2002ej}.  We quote the form of the
sum rules in 4D deconstructed theories in Table \ref{tab:sumrules}; those for the 5D theories correspond to the special case of setting $\Gt_4 = 0$ and $\Gt_{41} = 0$.

\subsection{
The Equivalence Theorem in 4D and 5D
}
\label{subsec41}

The Equivalence Theorem (ET) in 4D connects the
longitudinal gauge boson amplitude to the corresponding
would-be Nambu-Goldstone boson amplitude in any gauge theory with
spontaneous symmetry breaking\,\cite{He:1997zm}
,
\beq \label{eq:ET-4D} \ba{l} \dis
 T[W_{n_1L}^{a_1},W_{n_2L}^{a_2},\cdots,W_{n_rL}^{a_r}; \Phi_\alpha] ~=~
 T[i\pit_{n_1}^{a_1}, i\pit_{n_2}^{a_2}, \cdots,
   i\pit_{n_r}^{a_r}; \Phi_\alpha ] + B \,,
\\[4mm]
\dis
 B ~\equiv~ \sum_{\ell = 1}^r \(
 T[v_{n_1}^{a_1},\cdots,v_{n_{\ell}}^{a_{\ell}},
   i\pit_{n_{\ell+1}}^{a_{\ell+1}},\cdots , i\pit_{n_r}^{a_r}; \Phi_\alpha]
 +{\rm permutations}\)\,.
\ea
\eeq
The symbol $\Phi_\alpha$ denotes any other possible amputated external physical
fields, such as the transverse gauge boson $W_{nT}^a$. All external lines are directed inwards, and the $v^a_n$ are defined as
\beq
v_{n_i}^{a_i} \equiv v^\mu_{n_i} (p_i)\,W^{a_i}_{\mu n_i}\,,  \qquad\qquad  v^\mu_{n_i}(p_i) \equiv \epsilon^\mu_{Li} -\frac{p^\mu_i}{M_{n_i}} = {\cal O} \left(\frac{M_{n_i}}{E_i}\right) ,
\eeq
where $a$ is a gauge-group index, $\mu$ is a Lorentz index, $n$ labels the KK level, $i$ identifies each external state in the scattering process, and $p_i$ is the momentum of that state.
For simplicity we have dropped a multiplicative factor
$C=1+{\cal O}({\rm loop})$ associated with wavefunction renormalization of
the external Nambu-Goldstone fields on
the RHS, which will not affect our present analysis.\, Power
counting\,\cite{He:1997zm}
 shows that the longitudinal gauge boson
amplitude contains terms of ${\cal O}(E^4)$, ${\cal O}(E^2)$ and ${\cal O}(E^0)$, while
the Nambu-Goldstone boson amplitude only contains terms of ${\cal O}(E^2)$
(proportional to quartic coupling $\Gt_4$) and ${\cal O}(E^0)$. Therefore
the ET enforces an exact $E^4$-cancellation in the longitudinal
gauge boson amplitude, and further requires the $E^2$-terms of the
longitudinal amplitude to exactly equal the $E^2$-terms in the
Nambu-Goldstone amplitude (the first term on the RHS of the first
equation in (\ref{eq:ET-4D})). 

A general estimate of the $B$-term in
(\ref{eq:ET-4D}) gives\,\cite{He:1994br},
\beqa 
\label{eq:B} 
B &=&
   {\cal O}\!\(\!\f{M_j^2}{E_j^2}\!\)T[i\pit_{n_1}^{a_1},\cdots,i\pit_{n_r}^{a_r}; \Phi_\alpha]
 + {\cal O}\!\(\!\f{M_j}{E_j}\!\)T[W_{n_1T_j}^{a_{1}},
   i\pit_{n_2}^{a_{2}},\cdots,i\pit_{n_r}^{a_{r}}; \Phi_\alpha]\,.
\eeqa
For the two-body scattering processes in deconstructed models, 
the leading contributions to the amplitude
   $T[\widetilde{\pi}^{a_1}_{n_1},\widetilde{\pi}^{a_2}_{n_2},
      \widetilde{\pi}^{a_3}_{n_3},\widetilde{\pi}^{a_4}_{n_4}]$
scale like $E^2$, while the leading contribution to the mixed amplitude
   $T[W^{a_1}_{n_1T_j},\widetilde{\pi}^{a_2}_{n_2},
      \widetilde{\pi}^{a_3}_{n_3},\widetilde{\pi}^{a_4}_{n_4}]$ scales like $E^1$.
These contributions arise from the $\tilde{G}_4$ and $\tilde{G}_{41}$
couplings in Eqs. (\ref{eq:deft4},\ref{eq:deft41}) and 
from  this we deduce that, in any deconstructed theory,
\beqa
\label{eq:B-4D}
B &=& {\cal O}\!\(\Gt  E^0\) \,+\, {\cal O}\!\(M^2E^{-2}\) \,.
\eeqa
where $M$ is the mass of an external line.

 In the 5D continuum limit the would-be Nambu-Goldstone fields
 $\left\{\pit_n^{a}\right\}$ become the 5th component
 of the corresponding 5D gauge fields $\{W_n^{5a}\}$. So we find that
 the Kaluza-Klein Equivalence Theorem (KK-ET) for a compactified 5D theory with arbitrary geometry
 takes the following form\,\cite{He:2004zr,SekharChivukula:2001hz},
\beq \label{eq:ET-5D} \ba{l} \dis
 T[W_{n_1L}^{a_1},W_{n_2L}^{a_2},\cdots,W_{n_rL}^{a_r}; \Phi_\alpha] ~=~
 T[iW_{n_1}^{5a_1}, iW_{n_2}^{5a_2}, \cdots,
   iW_{n_r}^{5a_r} ; \Phi_\alpha] + B \,.
\ea
\eeq
Since there are no 3- and 4-$W_n^{5a}$ couplings in any continuum 5D Yang-Mills theory, 
the couplings $\Gt_4$ and $\Gt_{41}$ vanish in the continuum limit. In 5D,
since the LHS of Eq. (\ref{eq:ET-5D}) is purely longitudinal,  the $B$-term in (\ref{eq:B-4D}) reduces to
\beqa
\label{eq:B-5D}
B &=& {\cal O}\!\(M^2E^{-2}\) \,,
\eeqa
where, again, $M$ is the mass of an external line.
Therefore, using the longitudinal amplitude on the LHS of
(\ref{eq:ET-5D}) and the Nambu-Goldstone amplitude on its RHS, we can derive a new sum rule at ${\cal O}(E^0)$
valid for continuum 5D models which do not involve
the $B$-term.

We stress that the sum rules we are about to derive from the ET
 (\ref{eq:ET-4D}) or (\ref{eq:ET-5D})\,\cite{He:2004zr,SekharChivukula:2001hz}  do not
 depend on any particular boundary conditions: they hold for general
deconstructed theories and the corresponding 5D theories with
arbitrary 5D geometry.  For instance, it is straightforward to
verify that they all hold for the 5D $S^1/\mathbf{Z}_2$
model studied in \cite{SekharChivukula:2001hz}.

\subsection{Sum Rules from $LL \to LL$ Scattering}
\label{subsec4B}

\subsubsection{
Equivalence Theorem at Order $E^4$
}
\label{subsec42}

As discussed above, the ET  (\ref{eq:ET-4D}) enforces an exact $E^4$-cancellation
in the longitudinal scattering amplitude for either deconstructed
theories or continuum theories, i.e.,
\beqa
T_W[nn,mm]_{E^4} &=& 0 \,.
\eeqa
Based on the form of the $E^4$ terms in the amplitude of Eq. (\ref{eq:WL-E4}),
we obtain the following two sum rules at ${\cal O}(E^4)$,
\beqa
\label{eq:E4-sum1}
G_4^{nnmm} &=& \sum_k G_3^{nnk}G_3^{mmk} \,,
\\
\label{eq:E4-sum2}
G_4^{nnmm} &=& \sum_k \(G_3^{nmk}\)^2 \,.
\eeqa
For the special case of elastic scattering,
Eqs.\,(\ref{eq:E4-sum1})-(\ref{eq:E4-sum2})
reduce to a single sum rule,
\beqa
\label{eq:E4-sum-el}
G_4^{nnnn} &=& \sum_k \(G_3^{nnk}\)^2 \,,
\eeqa
which is of the same form as the one discussed in ref. \cite{Csaki:2003dt}.

Indeed, these sum rules directly follow from the fact that the mass-squared matrix for the gauge bosons of a spontaneously broken Yang-Mills theory is real and symmetric -- and is thus diagonalized by an orthogonal matrix.  We can explicitly verify them by  inserting the general definitions (\ref{eq:def3})-(\ref{eq:def4}) for the cubic and quartic 
gauge couplings into Eqs.\,(\ref{eq:E4-sum1})-(\ref{eq:E4-sum2}) and applying the orthonormality condition
$~\mathbf{R}\mathbf{R}^T=\mathbf{R}^T\mathbf{R}=I$. We will return to this point in Section \ref{WTI-CR-sec}.

\subsubsection{
Equivalence Theorem at Order $E^2$
}
\label{subsec43}

As discussed in Section \ref{subsec41},  the ET (\ref{eq:ET-4D}) enforces an equality between the gauge and Nambu-Goldstone scattering amplitudes at order $E^2$,
\beqa
\label{eq:ET-E2}
T_W[nn,mm]_{E^2} &=& \Tt_\pi [nn,mm]_{E^2} \,.
\eeqa
Applying this to Eqs. (\ref{eq:WL-E2}) and (\ref{eq:Tpi-E2}), we can deduce two independent conditions,
\beqa
\label{eq:E2-sum1}
2M_+^2G_4^{nnmm} + \sum_k\(\!G_3^{nmk}\!\)^2
\[\f{M_-^4}{M_k^2} - 3M_k^2\] &=& \f{4M_n^2M_m^2}{v^2}\Gt_4^{nnmm} \,,
\\[2mm]
\label{eq:E2-sum2}
\sum_k\[\(\!G_3^{nmk}\!\)^2 - G_3^{nnk}G_3^{mmk}\]\! M_k^2
&=&
\sum_k\(\!G_3^{nmk}\!\)^2\f{M_-^4}{M_k^2}   \,.
\eeqa
We note that the new sum rule (\ref{eq:E2-sum2}) does not depend on the 
quartic coupling $\Gt_4^{nnmm}$, so it is unchanged in the continuum limit where 
$\Gt_4^{nnmm} = 0$ holds. 
Substituting Eq. (\ref{eq:E2-sum2}) into Eq. (\ref{eq:E2-sum1}) and applying Eq. 
(\ref{eq:E4-sum2}), we obtain
\beqa
\label{eq:E2-sum1y}
\sum_k\[\(G_3^{nmk}\)^2\(M_+^2-M_k^2\) -
        \f{1}{2}G_3^{nnk}G_3^{mmk}M_k^2\]
&=&
\f{2M_n^2M_m^2}{v^2}\Gt_4^{nnmm} \,,
\eeqa
Taking the special case of elastic scattering ($n=m$),
we immediately see that the general sum rule  (\ref{eq:E2-sum1})
reduces to
 \beqa
 \label{eq:E2-sum-el}
  G_4^{nnnn} - \sum_k \(\!G_3^{nnk}\!\)^2 \f{\,3M_k^2\,}{\,4M_n^2\,}
 &=& \f{M_n^2}{v^2}\Gt_4^{nnnn}  \,,
 \eeqa
which is a generalization of  the continuum limit $\(\Gt_4^{nnnn}=0\)$ 
sum rule derived in \cite{Csaki:2003dt}.  
Our second sum rule (\ref{eq:E2-sum2}) at order $E^2$, on the other hand, 
becomes a tautology for $m=n$, implying that this new sum rule exists
only for inelastic channels.

\subsubsection{Equivalence Theorem at Order $E^0$
}

Using the sum rules derived above, we may further simplify the
longitudinal gauge boson amplitude
(\ref{eq:WL-E0}) at ${\cal O}(E^0)$, obtaining the form
\beqa
(T_1\!+\!T_2)_{E^0}^{~}
&\!\!\!=\!\!\!&
\f{E^0}{4M_n^2M_m^2}\!\left\{\!
\sum_k\(\!G_3^{nmk}\!\)^{\!2}\!\[\(\!M_k^4\!-\!\f{1}{3}M_+^4\!\)
        \!-\!2M_+^2\!\(\!M_k^2\!-\!\f{2}{3}M_+^2\!\)\]\!
        \f{\,3\!+\!c^2\,}{1\!-\!c}\, +
\right.
\nonumber
\\[2mm]
&&   \hspace*{18mm}
\sum_k\!\[\(\!G_3^{nmk}\!\)^2\!\!\!-\!G_3^{nnk}G_3^{mmk}\]\!\!
      \(M_k^4\!-\!2M_+^2M_k^2\)\! c
+G_4^{nnmm}M_-^4c -
\nonumber\\
&& \hspace*{18mm}  \left.
 \f{\,4\Gt_4^{nnmm}\,}{v^2}M_n^2M_m^2M_+^2(1-c)
\right\} .
 \label{eq:WL-E0x}
\eeqa
For convenience we also rewrite the Nambu-Goldstone boson amplitude
(\ref{eq:Tpi-E0}) as follows,
 \beqa
 \label{eq:Pi-E0x}
 (\Tt_1\!+\!\Tt_2)_{E^0}^{~} &\!\!\!=\!\!\!&
 \sum_k\left\{
 \(\!\Gt_3^{nmk}\!\)^{\!2}\f{\,3\!+\!c^2}{\,4(1\!-\!c)\,} +
 \[\(\!\Gt_3^{nmk}\!\)^{\!2}-\Gt_3^{nnk}\Gt_3^{mmk}
 \]\f{c}{4}\,
 \right\}
 \\
 && -\f{\,\Gt_4^{nnmm}}{v^2}M_+^2\(\f{1}{3}+c\)  .
 \nonumber
 \eeqa
 As discussed in Section \ref{subsec41}, these $E^0$ terms in the gauge and Nambu-Goldstone scattering amplitudes are related by the ET (\ref{eq:ET-4D}).   
 
 For models with vanishing quartic Nambu-Goldstone boson coupling
$\Gt_4^{nnmm}=0$ (such as 5D continuum theories), we find that the $B$-term is ${\cal O}(E^{-2})\,$ as in (\ref{eq:B-5D}), so the ET condition (\ref{eq:ET-E0}) reduces to,
\beqa
\label{eq:ET-E0b}
T_W[nn,mm]_{E^0} &=& \Tt_\pi [nn,mm]_{E^0}  \,,~~~~~~
\({\rm for}~\Gt_4=0\).
\eeqa
Applying this to Eqs. (\ref{eq:WL-E0x}) and (\ref{eq:Pi-E0x}) yields the sum rules
 \beqa
 \label{eq:E0-sum1}
 \sum_k\!\(\!G_3^{nmk}\!\)^{\!2}\!
 \[\f{\,M_k^2-M_+^2\,}{M_nM_m}\]^2
 &=&
 \sum_k\(\!\Gt_3^{nmk}\!\)^2 ,
 \\[3mm]
 \label{eq:E0-sum2}
 \hspace*{-3mm}
 G_4^{nnmm}M_-^4 \!+\! \sum_k\!\[\(\!G_3^{nmk}\!\)^{\!2}\!\!
 -\! G_3^{nnk}G_3^{mmk}\]\!\!\(M_k^4\!\!-\! 2M_+^2M_k^2\)
 &=&
 M_n^2M_m^2\!\sum_k\!\!\[\!\(\!\Gt_3^{nmk}\!\)^{\!2}\!\!
 -\! \Gt_3^{nnk}\Gt_3^{mmk}\!\]\! .
 \nonumber
 \\[-2mm]
 &&
 \eeqa
As we will see shortly, both continue to hold even
 for $\,\Gt \neq 0\,$.
 For the special case of elastic scattering ($n=m$), the sum
 rule (\ref{eq:E0-sum1}) reduces to,
 \beq
 \label{eq:E0-sum1-el}
 \sum_k \(\!G_3^{nnk}\!\)^{\!2}\[\f{M_k^2}{\,M_n^2\,} -2\]^2
= \sum_k \(\!\Gt_3^{nnk}\!\)^{\!2} .
 \eeq
 On the other hand, for $n=m$,
the sum rule (\ref{eq:E0-sum2}) is trivially satisfied,
 i.e. it exists only for the inelastic scattering.

In general 4D deconstructed theories with nonzero quartic Nambu-Goldstone boson
coupling $\Gt_4^{nnmm}\neq 0$, the $B$-term 
is actually of ${\cal O}(\Gt_4E^0)$ (cf. Eq.\,(\ref{eq:B-4D})).
This means the new sum rules at ${\cal O}(E^0)$, as imposed by the ET
(\ref{eq:ET-4D}),
\beqa
\label{eq:ET-E0}
T_W[nn,mm]_{E^0} &=& \Tt_\pi [nn,mm]_{E^0} + B_{E^0}  \,,
\eeqa
will involve the couplings $\Gt_4$ and $\Gt_{41}$, from Eq. (\ref{eq:deft4}) and (\ref{eq:deft41}).
Generally speaking, we can decompose the $B$-term according to its weak-isospin structure,
 \beqa
 B[nn,mm]
 & = &
 \ep^{abe}\ep^{cde}B_1 +
 \ep^{ace}\ep^{bde}B_2 +
 \ep^{ade}\ep^{bce}B_3
 \nonumber
 \\
 &=& \ep^{ace}\ep^{bde}(B_1+B_2) + \ep^{ade}\ep^{bce}(-B_1+B_3)
 \\
 &=&
 \d^{ab}\d^{cd}(B_2+B_3) +
 \d^{ac}\d^{bd}(B_1-B_3) +
 \d^{ad}\d^{bc}(-B_1-B_2)
 \nonumber
 \eeqa
 In this case, the term $B_{E^0}^{~}$
 consists of four diagrams involving the $\pit-\pit-\pit-W_k^\mu$ contact vertex
 (\ref{eq:Lint-piW}):
 \beqa
 \label{eq:B0-all}
 B^{~}[nn,mm]_{E^0}&=&
 T[v_{n\mu}(p_1)^{~}W_n^{a\mu},\,i\pit_n^b,\,-i\pit_m^c,\,-i\pit_m^d] +
 T[i\pit_n^a,\,v_{n\mu}(p_2)^{~}W_n^{b\mu},\,-i\pit_m^c,\,-i\pit_m^d] +
 \nonumber\\
 &&
 T[i\pit_n^a,\,i\pit_n^b,\,v_{m\mu}(p_3)^{~}W_m^{c\mu},\,-i\pit_m^d] +
 T[i\pit_n^a,\,i\pit_n^b,\,-i\pit_m^c,\,v_{m\mu}(p_4)^{~}W_m^{d\mu}] \,,
 \eeqa
 where each $v_{n_j}^\mu(p_j) \equiv   \eph_{Lj}^\mu -\f{p_j^\mu}{M_n}$ is computed to its leading order  at ${\cal O}(M_{n_j}/E)$.  We find
 {\small
 \beq
 B_{E^0}[nn,mm] =  \f{\,\Delta_{nm}\,}{3v} \[
 2\d^{ab}\d^{cd} - \d^{ac}\d^{bd}(1\!+\!3c) - \d^{ad}\d^{bc}(1\!-\!3c)
 \] + O\({M_{n,m}^2}/{E^2}\) \,,
 \label{eq:B0-sum}
 \eeq
where
\beq
 \dis
 \Delta_{nm} \equiv
 M_n\Gt_{41}^{mmnn} + M_m\Gt_{41}^{nnmm} \,.
 \label{eq:Delta}
 \eeq
 So we arrive at
 \beq
 \label{eq:B1+B2-E0}
 (B_1+B_2)_{E^0}  ~=~ +\f{\,\Delta_{nm}\,}{3v}(1-3c) \,.
 \eeq

 Finally we substitute (\ref{eq:WL-E0x}), (\ref{eq:Pi-E0x})
 and (\ref{eq:B1+B2-E0}) into the ET identity (\ref{eq:ET-E0}), in order to find the sum rules that apply for 4D deconstructed models.   The ET identity (\ref{eq:ET-E0}) requires the total coefficients of
 $\f{\,3+c^2}{1-c}$ and $c$, and the $c$-independent constant term to vanish separately.  The first two conditions yield the same sum rules, Eqs. (\ref{eq:E0-sum1}) and (\ref{eq:E0-sum2}), as we derived in the continuum case. The coefficient of the c-independent constant term (arising from the quartic
 couplings $\Gt_4^{nnmm}$ and $\Gt_{41}^{nnmm}$) has the form,
  \beqa
  \label{eq:4GB-term-E0}
 \[\f{\,2\Gt_4^{nnmm}\,}{3v^2}M_+^2 +
   \f{\Delta_{nm}}{3v}\] \,.
 \eeqa
Requiring this to vanish gives rise to a new constraint at ${\cal O}(E^0)$.  Inserting the definition (\ref{eq:Delta}) puts this constraint in the form 
 of a relationship between the two quartic couplings in the Nambu-Goldstone sector,
  \beqa
 \label{eq:E0-sum3f}
 M_n\Gt_{41}^{mmnn} + M_m\Gt_{41}^{nnmm} &=&
 - \Gt_4^{nnmm}\f{2M_+^2}{v} \,.
 \eeqa
 %


\subsection{Sum Rules from $LL \to LT$ Scattering}

In addition to the purely longitudinal scattering processes we have investigated thus far, there are 
  additional scattering amplitudes involving a mix of longitudinal and transverse gauge bosons, where the ET (\ref{eq:ET-4D})
 can be applied to yield further constraints on the couplings and masses.
Of particular interest is $W_{nT}^aW_{nL}^b \to W_{mL}^cW_{mL}^d$ (and processes related by crossing), since the leading term in the Nambu-Goldstone amplitude (see Sec. III C) involves only the coupling $\Gt_{41}^{nnmm}$ which we have just constrained in Eq. (\ref{eq:E0-sum3f}). This suggests that applying the ET to $W_{nT}^aW_{nL}^b \to W_{mL}^cW_{mL}^d$ will yield a new sum rule involving $\Gt_{41}^{nnmm}$ and the gauge couplings $G_4^{nnmm}$ and
 $G_3^{nmk}$.  We will show that the new sum rule,  together with (\ref{eq:E0-sum3f}), fully determines
 $\Gt_{41}^{nnmm}$ in terms of other couplings and masses.

\subsubsection{ Equivalence Theorem at ${\cal O}(E^3)$ }

 Because the Nambu-Goldstone boson amplitude vanishes at ${\cal O}(E^3)$,
 i.e.,  $\,\Tt[W_{nT}\pit_n,\pit_m\pit_m]_{E^3}=0\,$ (cf. Sec.\,III C)\,, the ET
 constrains the sum of $E^3$-terms (\ref{eq:TLLL-T1+T2-E3}) to equal zero,
 which results in the now-familiar sum rules, (\ref{eq:E4-sum1})-(\ref{eq:E4-sum2}) previously found for $LL\to LL$ scattering at order $E^4$. This is not unexpected,
 because the $E^3$-cancellation involves only the pure Yang-Mills
 interactions, not the gauge boson mass-term
 (or the Nambu-Goldstone interactions on the RHS of the ET).

\subsubsection{ Equivalence Theorem at ${\cal O}(E^1)$ }

 For the scattering
 $W_{nT}^aW_{nL}^b\to W_{mL}^cW_{mL}^d$ ($W_{nT}^a\pit_n^b\to
 \pit_m^c\pit_m^d$), we find that applying the ET to Eqs. (\ref{eq:TLLL-T1+T2-E1}) and
 (\ref{eq:Tpipipi-T1+T2-E1}) imposes a condition  
 \beq
 \label{eq:TLLL-sum-E1}
  \f{2M_nM_m^2}{v}\Gt_{41}^{mmnn} =
  -3M_+^2G_4^{nnmm} + \sum_k G_3^{nnk}G_3^{mmk}2(2M_n^2 \!-\! M_k^2)
  -\sum_k\(\!G_3^{nmk}\!\)^2\!\(
  M_k^2 -(M_n^2 \!+\! 5M_m^2) + \f{M_-^4}{M_k^2}
  \).
 \eeq
 Similarly, for the scattering
 $W_{nL}^aW_{nL}^b\to W_{mL}^cW_{mT}^d$ as in (\ref{eq:TLLL-T1+T2-E1}) and the scattering 
 $\pit_n^a\pit_n^b\to
 \pit_m^c W_{mT}^d$ as in 
 (\ref{eq:Tpipipi-T1+T2-E1}) with $n \leftrightarrow m$, we deduce, by explicit calculation,
 \beq
 \label{eq:LLLT-sum-E1}
  \f{2M_n^2M_m}{v}\Gt_{41}^{nnmm} =
  -3M_+^2G_4^{nnmm} + \sum_k G_3^{nnk}G_3^{mmk}2(2M_m^2 \!-\! M_k^2)
  -\sum_k\(\!G_3^{nmk}\!\)^2\!\(
  M_k^2 -(M_m^2 \!+\! 5M_n^2) + \f{M_-^4}{M_k^2}
  \).
 \eeq
  By taking the difference between (\ref{eq:TLLL-sum-E1}) and
  (\ref{eq:LLLT-sum-E1}), we derive a condition,
  \beq
  \f{2M_nM_m}{v}\(M_n\Gt_{41}^{nnmm} - M_m\Gt_{41}^{mmnn}\)
  = 
  4M_-^2\sum_k\[\(\!G_3^{nmk}\!\)^2 - G_3^{nnk}G_3^{mmk}\]\, .
\eeq
The sum rules (\ref{eq:E4-sum1}) and (\ref{eq:E4-sum2}) tell us that the RHS is zero, yielding 
  \beqa
  \label{eq:G41-sum-E1-}
   M_n\Gt_{41}^{nnmm} \!&=&\! M_m\Gt_{41}^{mmnn}  \,.
  \eeqa
On the other hand, if we take the sum of Eqs. (\ref{eq:TLLL-sum-E1}) and
  (\ref{eq:LLLT-sum-E1}), we derive another condition,
  \beqa
  \f{M_nM_m}{v}\(M_n\Gt_{41}^{nnmm} + M_m\Gt_{41}^{mmnn}\)
  &\!\!\!=\!\!\!& -\f{\,4M_n^2M_m^2\,}{v^2}\Gt_4^{nnmm}  \,,
  \eeqa
  which simplifies as 
  \beqa
  \label{eq:G41-sum-E1+}
  M_n\Gt_{41}^{nnmm} + M_m\Gt_{41}^{mmnn}
  &\!\!\!=\!\!\!&
  -\f{\,4M_nM_m\,}{v}\Gt_4^{nnmm}  \,.
  \eeqa
  Substituting (\ref{eq:G41-sum-E1-}) into
  (\ref{eq:G41-sum-E1+}) yields two expressions for
  $\Gt_{41}$,
  \beq
  \label{eq:G41a}
  \dis
  \Gt_{41}^{nnmm} ~=~ -\f{2M_m}{v}\Gt_4^{nnmm} \,,
  ~~~~~~
  \Gt_{41}^{mmnn} ~=~ -\f{2M_n}{v}\Gt_4^{nnmm} \,.
  \eeq
  Similarly, if we substitute (\ref{eq:G41-sum-E1-}) into
  (\ref{eq:E0-sum3f}), we again obtain the relationships (\ref{eq:G41a}).
  Note that the 4-Nambu-Goldstone coupling $\Gt_4^{nnmm}$ is symmetric
  under indices $(n,m)$, i.e., $\,\Gt_4^{nnmm}=\Gt_4^{mmnn}\,$.\,
  
\subsection{Sum Rules from $LL \to TT$ Scattering}

For the scattering process
 $W_{nL}W_{nL}\to W_{mT}W_{mT}~$ ($\pi_n\pi_n\to W_{mT}W_{mT}$),
 the Equivalence Theorem (ET) takes the form,
 \beqa
 \label{eq:ET}
 T_W [nn,mm] &=& T_\pi [nn,mm] \,+\, B \,,
  \\
  B &\equiv& \T [v_\mu^{} W_n^{a\mu},i\pi_n^b;W_{mT}^c,W_{mT}^d]
          +\T [i\pi_n^a,v_\mu^{} W_n^{b\mu};W_{mT}^c,W_{mT}^d] \,.
 \eeqa
 For the current case, it is straightforward for us to explicitly count the
 $E$-power dependence of the $B$-term under the $M_{W_n}/E$
 expansion, and we find
 \beqa
 B &=& {\cal O}\!\(E^{-2}\) \,,
 \eeqa
 where the absence of a 4-point 
 \,$\pi$-$W$-$W$-$W$}\, contact vertex is crucial.  Hence, the $B$ term will be negligible in this case.

\subsubsection{Equivalence Theorem at ${\cal{O}}(E^2)$}

Because the scattering amplitude involving Nambu-Goldstone bosons 
has no order $E^2$ term, the coefficient of $E^2$ in the $W_{nL}W_{nL}\to W_{mT}W_{mT}~$ 
amplitude must vanish; this holds for Eqs. (\ref{eq:LL-TT-E2}) and (\ref{eq:PiPi-TT}) in which polarization vectors  $\eph_{3T}^{}$ and $\eph_{4T}^{}$ were used and also for Eqs.  (\ref{eq:LL-TT'-E2}) and (\ref{eq:PiPi-TT'}) where polarizations 
$\eph_{3T'}^{}$ and $\eph_{4T'}^{}$ were used instead.  The results duplicate
the previously derived sum rules (\ref{eq:E4-sum1})-(\ref{eq:E4-sum2}).

\subsubsection{Equivalence Theorem at ${\cal{O}}(E^0)$}

We will apply the ET, in turn, to the amplitudes calculated using polarization vectors 
$\eph_{3T}^{}$ and $\eph_{4T}^{}$ and those using $\eph_{3T'}^{}$ and $\eph_{4T'}^{}$.  In the first case, 
the ET requires that the right-hand-sides of Eqs. (\ref{eq:LL-TT-E0}) and  (\ref{eq:PiPi-TT}) be equal.   From the terms that are independent of scattering angle, we find the sum rule,
\beqa
 \label{eq:SR-E0-TT-1}
 G_4^{nnmm} +{\dis\sum_k}\! \f{(G_3^{nmk})^2}{2}
 \(\!\f{\Mb_-^4}{\Mb_k^2}-\Mb_k^2\)
  &=&
 \Gt_{42}^{nnmm} + \sum_k\! \f{1}{2}(\Gt_3^{nkm})^2 ,
\eeqa
while from the terms proportional to $c$ we obtain,
 \beqa
 \label{eq:SR-E0-TT-2}
 \hspace*{-10mm}
 {\dis\sum_k}\!\[\! G_3^{nnk}G_3^{mmk}\!(\Mb_k^2\!-\!2\Mb_m^2)\!
 + (G_3^{nmk})^2\!
  \(\f{\Mb_-^4}{\Mb_k^2}\!+\!\Mb_k^2\!\)\!\]
 &=&
 \sum_k\!\[\! (\Gt_3^{nkm})^2 \!-\!\Gt_3^{nnk}G_3^{mmk}\!\]\!.
 \eeqa
Note that the Nambu-Goldstone coupling
 $\Gt_3^{nkm}$ differs from $\Gt_3^{nmk}$ when $k\neq m$ because here
 the first two indices denote the Nambu-Goldstone bosons and the third
 index always denotes the gauge boson by definition.
 This is contrast to the triple gauge coupling $G_3^{nmk}$
 which is fully symmetric under exchange of its three indices.

In the second case, we obtain, instead, the sum rules,
 \beqa
 \label{eq:SR-E0-TT'-1}
 G_4^{nnmm} +
 {\dis\sum_k}(G_3^{nmk})^2\(\Mb_-^2 \!-\! \Mb_k^2\)
 &=& \Gt_{42}^{nnmm}  \,,
 \\
 \label{eq:SR-E0-TT'-2}
 \hspace*{-10mm}
 {\dis\sum_k}\!
 \[  G_3^{nnk}G_3^{mmk}(2\Mb_m^2\!-\!\Mb_k^2)
   - 2(G_3^{nmk})^2\Mb_-^2 \]
 &=&
 \sum_k \Gt_3^{nnk}G_3^{mmk} ~,
 \eeqa
which contain additional information.

It turns out that these four new sum rules are not all independent of one another.
The sum of (\ref{eq:SR-E0-TT-1}) with (\ref{eq:SR-E0-TT'-1}) and the difference between 
(\ref{eq:SR-E0-TT-2}) and (\ref{eq:SR-E0-TT'-2}) both yield the same equation:
\beqa
 \label{eq:SR-E0-LLTT-1in}
 \sum_k (\Gt_3^{nkm})^2
 &=&
 \sum_k (G_3^{nmk})^2\!\(\!\Mb_k - \f{\Mb_-^2}{\Mb_k}\!\)^{\!\!2},
 \eeqa
The complementary combinations, however, result in two independent equations,
\beqa
 \label{eq:SR-E0-LLTT-2in}
 G_4^{nnmm} &=&
 \sum_k\hf\!\[  \Gt_3^{nnk}G_3^{mmk}
                 +G_3^{nnk}G_3^{mmk}\Mb_k^2 \]\!,
 \\
 \label{eq:SR-E0-LLTT-3in}
 \Gt_{42}^{nnmm} &=& G_4^{nnmm}\Mb_m^2 -\sum_k(G_3^{nmk})^2\Mb_k^2\,.
 \eeqa
Thus, the $LL \to TT$ scattering yields a total of three new relations among the 
gauge and Nambu-Goldstone couplings.  

For the special case of elastic scattering, the three sum rules reduce to,
 \beqa
 \label{eq:SR-E0-LLTT-1e}
 \sum_k (\Gt_3^{nkn})^2 &=&
 \sum_k (G_3^{nnk})^2\(\f{M_k}{M_n}\)^2 \,,
 \\
 \label{eq:SR-E0-LLTT-2e}
 G_4^{nnnn}  &=& \sum_k
 \hf \[(\Gt_3^{nkn})^2 + G_3^{nnk}\Gt_3^{nnk}\] \,,
 \\
 \label{eq:SR-E0-LLTT-3e}
 \Gt_{42}^{nnnn} &=& G_4^{nnnn} - \sum _k (\Gt_3^{nkn})^2
 ~=~ \sum_k\[(G_3^{nkn})^2-(\Gt_3^{nkn})^2\]~.
 \eeqa
 %

\subsection{Sum Rules from $LT \to TT$ Scattering}

Applying the Equivalence Theorem to the amplitudes in section \ref{sub:LTTT} simply 
constrains the ${\cal O}(E)$ terms from the gauge scattering process to vanish.  This yields, 
once again, the sum rules (\ref{eq:E4-sum1})-(\ref{eq:E4-sum2}).
 
 \ \ 
 
 All of the sum rules derived in this section are summarized in Table \ref{tab:sumrules}.

\section{Implications of Gauge Invariance 
}
\label{Rxi-sec}

\subsection{$W_{n}^aW_{n}^b\to W_{m}^cW_{m}^d$ in $R_\xi$-Gauge}

The calculation of this process in $R_\xi$ gauge differs in two major respects from the unitary
 gauge calculation.  First,  the $t(u)$-channel $W_k$-exchange contains the
 propagator
 \beqa
 \label{eq:Wk-Rxi2}
 -i\[\f{g^{\mu\nu}}{p^2-M_k^2}+\f{(\xi-1)\,p^\mu p^\nu}{(p^2-M_k^2)(p^2-\xi M_k^2)}\]
 \eeqa
 rather than the unitary gauge propagator
  \beqa
 \label{eq:Wk-Uni}
 -i\f{~g^{\mu\nu}\! - p^\mu p^\nu/M_k^2~}{p^2-M_k^2} \,,
 \eeqa
which is the $\xi\to\infinity$ limit of (\ref{eq:Wk-Rxi2}).  Second, there are additional diagrams arising from $(s,t,u)$-channel exchange of the Goldstone mode $\pit_k$ with propagator
 \beqa
 \f{i}{\,p^2-\xi M_k^2\,} \,.
 \eeqa
These diagrams involve the $VV\pi$ vertices, whose coupling $\Gt_{31}^{nmk}$ (\ref{eq:deft31}) is antisymmetric under $n \leftrightarrow m$.

Let us start by considering the s-channel diagrams.  For $s$-channel $W_k$-exchange, the $p^\mu p^\nu$ term from the $W_k$-propagator vanishes identically for the  $nn\to mm$
 scattering due to the on-shell conditions $k_1^2=k_2^2=M_n^2$ and $k_3^2=k_4^2=M_m^2$. Moreover, the contribution of  the $s$-channel $\pit_k$-exchange diagram to $nn\to mm$ scattering also vanishes because \,$\Gt_{31}^{nnk}=\Gt_{31}^{mmk}=0$\,.  Hence, the only $s$-channel contribution to $WW$ scattering is that from the $g^{\mu\nu}$ term arising from $W_k$ exchange -- exactly as in unitary gauge.  Note that the $\xi$-independence in the $s$-channel $WW$-amplitude is
 automatically realized here, for arbitrary polarizations of the external gauge bosons and without any additional condition on the couplings and masses.  

Next, we analyze $t$-channel $W_k$-exchange in the $R_\xi$-gauge,
 \beq
 \label{eq:Tt-Rxi}
 \ba{ll}
 \hspace*{-3mm}
 \TT_t^R[W_k] &\!\!\!= \dis\ep^{ace}\ep^{bde}~\sum_k (G_3^{nmk})^2
 \[\f{~g^{\si\si'}\,}{t-M_k^2}
  +\f{~(\xi-1)(p_1 - p_3)^\si\!(p_1 - p_3)^{\si'}\,}{(t-M_k^2)(t-\xi M_k^2)}
 \]
 \Tb_{t,\si\si'}^k
 \,,
 \\[5mm]
 \hspace*{-3mm}
 \Tb_{t,\si\si'}^k &\!\!\! \equiv \dis
 \[(\eph_1^{}\!\cdot\!\eph_3^{})\,(p_1^{}\!+\!p_3^{})
    \!-\!2(p_3^{}\!\!\cdot\!\eph_1^{})\,\eph_3^{}
    \!-\!2(p_1^{}\!\!\cdot\!\eph_3^{})\,\eph_1^{}
 \]_\si
 \[ (\eph_2^{}\!\cdot\!\eph_4^{})\,(p_2^{}\!+\!p_4^{})
    \!-\!2(p_4^{}\!\!\cdot\!\eph_2^{})\,\eph_4^{}
    \!-\!2(p_2^{}\!\!\cdot\!\eph_4^{})\,\eph_2^{}\]_{\si'}\,.
 \ea
 \eeq
Since the $g^{\mu\nu}$ terms of the $W_k$ propagator are $\xi$-independent, we will focus on the $p^\mu p^\nu$ term.  We can compute the $(p_1-p_3)^\si\!(p_1-p_3)^{\si'}$ term for a given $k$ as
 \beqa
 \Delta_R
 & \equiv &  \f{~(\xi-1)(p_1-p_3)^\si\!(p_1-p_3)^{\si'}\,}{(t-M_k^2)(t-\xi M_k^2)}
 \Tb_{t,\si\si'}^k
 \nonumber\\
 &=&
 (\eph_1^{}\!\cdot\!\eph_3^{})(\eph_2^{}\!\cdot\!\eph_4^{})
 \f{\,(\xi -1)(p_3^2-p_1^2)(p_4^2-p_2^2)\,}{(t-M_k^2)(t-\xi M_k^2)}
 \nonumber\\
 &=&
 (\eph_1^{}\!\cdot\!\eph_3^{})(\eph_2^{}\!\cdot\!\eph_4^{})
 \f{M_-^4}{\,t-M_k^2\,}\,
 \f{\,\xi -1\,}{\,t-\xi M_k^2\,}
 \,.
 \eeqa
Likewise,  the $t$-channel $(p_1-p_3)^\si\!(p_1-p_3)^{\si'}$ term in unitary gauge is
 \beq
 \Delta_U
 \equiv   \f{\,-(p_1-p_3)^\si\!(p_1-p_3)^{\si'}\!\!/M_k^2\,}{\,t-M_k^2\,}
 \Tb_{t,\si\si'}^k
 = (\eph_1^{}\!\cdot\!\eph_3^{})(\eph_2^{}\!\cdot\!\eph_4^{})
 \f{M_-^4}{\,t-M_k^2\,}\,\[-\f{1}{M_k^2}\],
 \eeq
 so that the contributions from $W_k$ exchange in the two gauges differ by,
 \beq
 \Delta_R - \Delta_U
 =
 (\eph_1^{}\!\cdot\!\eph_3^{})(\eph_2^{}\!\cdot\!\eph_4^{})\f{M_-^4}{\,t-M_k^2\,}
 \[ \f{\,t-M_k^2\,}{\,M_k^2(t-\xi M_k^2)\,} \]
 =
 (\eph_1^{}\!\cdot\!\eph_3^{})(\eph_2^{}\!\cdot\!\eph_4^{})
 \, \f{\,M_-^4\,}{\,M_k^2(t-\xi M_k^2)\,} \,.
 \eeq
 Hence, the complete  $t$-channel amplitudes from $W_k$ exchange differ in the two gauges by the amount
 \beqa
 \label{eq:RUt-Wk-diff}
 \TT_t^R[W_k] - \TT_t^U[W_k] &=& \dis\ep^{ace}\ep^{bde}~ \sum_k ( \Delta_R - \Delta_U) (G_3^{nmk})^2
 \\
 & = & \dis\ep^{ace}\ep^{bde}~
 (\eph_1^{}\!\cdot\!\eph_3^{})(\eph_2^{}\!\cdot\!\eph_4^{})\sum_k
 \f{\,M_-^4(G_3^{nmk})^2\,}{\,M_k^2(t-\xi M_k^2)\,} \,.
 \eeqa
 Similarly for the $u$-channel $W_k$-exchange we have
 \beqa
 \label{eq:RUu-Wk-diff}
 \TT_u^R[W_k] - \TT_u^U[W_k] & = & \dis\ep^{ade}\ep^{bce}~
 (\eph_1^{}\!\cdot\!\eph_4^{})(\eph_2^{}\!\cdot\!\eph_3^{})\sum_k
 \f{\,M_-^4(G_3^{nmk})^2\,}{\,M_k^2(u-\xi M_k^2)\,} \,.
 \eeqa

Goldstone $\pit_k$-exchange contributes only in $R_\xi$ gauge;  in the $t$-channel and $u$-channel, we thus obtain, respectively,
 \beqa
 \label{eq:RUt-pik-diff}
 \TT_t^R[\pit_k] - \TT_t^U[\pit_k] & = & \dis\ep^{ace}\ep^{bde}~
 (\eph_1^{}\!\cdot\!\eph_3^{})(\eph_2^{}\!\cdot\!\eph_4^{})\sum_k
 \f{\,-(M_m + M_n)^2(\Gt_{31}^{nmk})^2\,}{\,t-\xi M_k^2\,} \,,
 \\
 \label{eq:RUu-pik-diff}
 \TT_u^R[\pit_k] - \TT_u^U[\pit_k] & = & \dis\ep^{ade}\ep^{bce}~
 (\eph_1^{}\!\cdot\!\eph_4^{})(\eph_2^{}\!\cdot\!\eph_3^{})\sum_k
 \f{\,-(M_m+M_n)^2(\Gt_{31}^{nmk})^2\,}{\,u-\xi M_k^2\,} \,.
 \eeqa

 It is evident that gauge-invariance ($\xi$-independence) of the physical $WW$-scattering
 amplitude must enforce an {\it exact cancellation} of the unphysical $\xi$-dependent mass-poles
 ($\dis 1/(t^2-\xi M_k^2)$ and $\dis 1/(u^2-\xi M_k^2)$)
 between the $W_k$-exchange and $\pit_k$-exchange contributions.
 From (\ref{eq:RUt-Wk-diff})-(\ref{eq:RUt-pik-diff}) and
 (\ref{eq:RUu-Wk-diff})-(\ref{eq:RUu-pik-diff}), we thus deduce the
 exact $\xi$-independence conditions,
 \beqa
 \sum_k
 \[\f{\,M_-^4(G_3^{nmk})^2\,}{\,M_k^2\,} - (M_m+M_n)^2(\Gt_{31}^{nmk})^2\]
 \f{1}{\,t-\xi M_k^2\,} &=& 0 \,,
 \\
  \sum_k
 \[\f{\,M_-^4(G_3^{nmk})^2\,}{\,M_k^2\,} - (M_m+M_n)^2(\Gt_{31}^{nmk})^2\]
 \f{1}{\,u-\xi M_k^2\,} &=& 0 \,,
 \eeqa
 from which we derive a  {\it new condition} for exact
 gauge-invariance,
 \beqa
 \label{eq:xi-cond-G3x}
 (\Gt_{31}^{nmk} M_k)^2 &=& ( G_3^{nmk} (M_n - M_m))^2 \,.
 \eeqa
 We note that both $\Gt_{31}^{nmk}$ and $M_n-M_m$ are
 anti-symmetric under exchange \,$n \leftrightarrow m$\,,\,
 while $G_3^{nmk}$ is symmetric with respect to indices $(n,m,k)$.
 We stress that our derivation of the $\xi$-independence condition
 (\ref{eq:xi-cond-G3x}) is completely general: valid for the gauge-boson amplitude with
 arbitrary polarizations of the external gauge boson states and independent of any
 detail of the model (either 4d-deconstruction or
 5d-compactification).

\subsection{$\pit_{n}^a\pit_{n}^b\to \pit_{m}^c\pit_{m}^d$ in $R_\xi$-Gauge}

Having verified that the amplitude for $WW$ scattering is gauge invariant, and therefore independent of $\xi$, we should now examine the $\xi$-dependent contributions to the process 
$\pit_{n}^a\pit_{n}^b\to \pit_{m}^c\pit_{m}^d$ in $R_\xi$ gauge.  These could potentially affect the sum rules derived earlier using the ET, where the $WW$-amplitudes were computed in unitary gauge
and the $\widetilde{\pi}\widetilde{\pi}$-amplitudes were derived in 't Hooft-Feynman
gauge.

 Let us consider the $(\xi\!-\!1)$-contributions of the $W_k$-propagator
 to the $s$, $t$ and $u$ channels, in turn.  The $s$-channel is the most straightforward to deal with.
 Due to the on-shell conditions $k_1^2=k_2^2=M_n^2$ and $k_3^2=k_4^2=M_m^2$, one may readily show that the $(\xi\!-\!1)p^\mu p^\nu$ term exactly vanishes in the
 $s$-channel.  However, for the $t$-channel, the $(\xi-1)p^\mu p^\nu$ term
 no longer vanishes and, instead, gives the following contribution,
 \beqa
 && \ep^{ace}\ep^{bde}\sum_k\f{1}{4} (\Gt_3^{nmk})^2
 (\xi-1)\f{[(p_1\!+\!p_3)\!\cdot\! (p_1\!-\!p_3)]
           [(p_2\!+\!p_4)\!\cdot\! (p_2\!-\!p_4)]}{(t-M_k^2)(t-\xi M_k^2)}
 \nonumber \\
 &=&
 \ep^{ace}\ep^{bde}\sum_k\f{1}{4} (\Gt_3^{nmk})^2
 (\xi-1)\f{(M_m^2\!-\!M_n^2)^2}{(t-M_k^2)(t-\xi M_k^2)}
 ~=~ O(E^{-4}) \,.
 \eeqa
 Similarly we can deduce the $u$-channel contribution from the
 $(\xi\!-\!1)p^\mu p^\nu$ term of the $W_k$-propagator,
 \beqa
 \ep^{ade}\ep^{bce}\sum_k\f{1}{4} (\Gt_3^{nmk})^2
 (\xi-1)\f{(M_m^2\!-\!M_n^2)^2}{(u-M_k^2)(u-\xi M_k^2)}
 ~=~ O(E^{-4}) \,.
 \eeqa
 The above proves that the $\xi$-dependent terms arising from the $W_k$-propagator actually vanish up to $O(1/E^3)$. Hence, these $\xi$-dependent terms will not affect  the sum rules  derived in Sec. \ref{subsec4B} via the ET at $O(E^\ell)$ ($\ell =4,3,2,1,0$).

\section{Deconstruction Identities in  Higgsless Models
}
\label{WTI-CR-sec}

This section presents an alternative way of deriving relations among the couplings and boson masses in Higgsless models.  We start by deriving completeness relations based explicitly on the orthogonality of the gauge boson diagonalization matrix $\bbR$ and the Nambu-Goldstone boson diagonalization matrix $\tilde{\bbR}$.  Then, we derive a set of Ward-Takahashi identities relating each coupling involving Nambu-Goldstone modes to couplings of gauge modes only.  

Several results presented here are specific to linear moose models and their 5D continuum limits because their derivation draws directly upon the form of the mixed gauge/Nambu-Goldstone couplings in Eqs. (\ref{eq:deft41}) - (\ref{eq:deft3}).  Within this class of models, however, the relations derived in this section form a generalization of the ET sum rules obtained in the previous section because they are not restricted to $nn\to mm$ processes and allow all of the coupling indices to vary independently.  
We will discuss more general applications of this method in Section \ref{sec:concl}.

\subsection{Completeness Relations}

Recall from Section \ref{sec:higgsless} that the gauge boson and Nambu-Goldstone boson mass matrices are  given, respectively,  by
\begin{equation}
  \bbM_W^2 = \bbQ^T \bbQ, \qquad\qquad\qquad   \tilde{\bbM}_W^2 = \bbQ \bbQ^T
\end{equation}
where $\bbQ$ may be written for linear moose models as (Eq. (\ref{eq:Q})),
\begin{equation}
  \bbQ = \frac{1}{2} \left(
    \begin{array}{ccccc}
      g_0 f_1 & - g_1 f_1 &           &             &
    \\
              & g_1 f_2   & - g_2 f_2 &             &
    \\
              &           &   \ddots  & \ddots      &
    \\
              &           &           & g_{N-1} f_N & -g_N f_N
    \\
              &           &           &             & g_N f_{N+1}
    \end{array}
  \right) .
\end{equation}
We likewise recall that the matrix $\bbQ$ is diagonalized by orthogonal matrices $\bbR$ and
$\tilde{\bbR}$, 
\begin{equation}
  \bbM_W^{\rm diag} = \bbR^T \bbQ^T \tilde{\bbR} 
                  = \tilde{\bbR}^T \bbQ \bbR, 
\label{eq:diagQ}
\end{equation}
where the matrix $\bbR$ describes the KK gauge-boson mass-eigenstate
``wave-functions'' in the deconstructed bulk, while $\tilde{\bbR}$ describes
the KK NG-boson mass-eigenstate ``wave-functions''.

We now use this information to derive some component-level expressions
that will be of use in our analysis.  From Eq.(\ref{eq:diagQ}) we see
\begin{equation}
  \tilde{\bbR} \bbM_W^{\rm diag} = \bbQ \bbR ,
  \label{eq:rtmqr}
\end{equation}
which implies
\begin{eqnarray}
  \tilde{\bbR}_{j,k} M_k &=& (\bbQ \bbR)_{j,k}
  \nonumber\\
  &=& \frac{1}{2}(g_{j-1} f_j \bbR_{j-1,k} - g_j f_j \bbR_{j,k}).
\label{eq:WT1}
\end{eqnarray}
Note that Eq.(\ref{eq:WT1}) relates the KK NG-boson wave-functions to 
the KK gauge-boson wave-functions.
It will be used in the derivation of Ward-Takahashi identities later.
In a similar manner, from Eq.(\ref{eq:five}), we see
\begin{equation}
  \bbR (\bbM_W^{\rm diag})^2 = \bbM_W^2 \bbR,
\end{equation}
and thus
\begin{eqnarray}
  \bbR_{j,n} M_n^2 &=& (\bbM_W^2 \bbR)_{j,n}
  \nonumber\\
  &=&\dfrac{1}{4} g_j^2 (f_j^2 + f_{j+1}^2) \bbR_{j,n}
    -\dfrac{1}{4} g_{j-1} g_j f_j^2 \bbR_{j-1,n}
    -\dfrac{1}{4} g_{j} g_{j+1} f_{j+1}^2 \bbR_{j+1,n} .
\label{eq:eigenmass-rel}
\end{eqnarray}
Because $\bbR$ is an orthogonal matrix, we may write
\begin{equation}
  \delta_{j,j'} = (\bbR \bbR^{T})_{j,j'} = \sum_{k} \bbR_{j,k} \bbR_{j',k}\,,
\label{eq:completeness1}
\end{equation}
which may be understood as a completeness
relation for the KK gauge-boson wave-function.
A similar relation hold for KK NG-boson wave-function,
$\tilde{\bbR}$, 
\begin{equation}
  \delta_{j,j'} = (\tilde{\bbR} \tilde{\bbR}^{T})_{j,j'} = \sum_{k} \tilde{\bbR}_{j,k} \tilde{\bbR}_{j',k}.
\label{eq:completeness2}
\end{equation}
%

\subsection{Identities from completeness relations}

We are now ready to derive identities among the various 3-point and 4-point couplings discussed in Section \ref{sec:higgsless}.  We first consider the product of three-point gauge-boson couplings:
\begin{equation}
  \sum_i G_{3}^{nmi} G_{3}^{\ell ki}
  = \sum_i \sum_{j,j'} g_j g_{j'} \bbR_{j, n}\bbR_{j, m} \bbR_{j, i}
    \bbR_{j', \ell}\bbR_{j', k} \bbR_{j', i} .
\end{equation}
Using the completeness relation Eq.(\ref{eq:completeness1}) resolves the
sums over $i$ and $j'$ on the RHS, yielding 
\begin{equation}
  \sum_i G_{3}^{nmi} G_{3}^{\ell ki}
  = \sum_{j} g_j^2 
    \bbR_{j, n}\bbR_{j, m} \bbR_{j,\ell}\bbR_{j, k}.
\end{equation}
Comparing this with the definition of $G_4$ in Eq.(\ref{eq:def4}) we
obtain an identity,
\begin{equation}
  \sum_i G_{3}^{nmi} G_{3}^{\ell ki}
  = G_4^{nm\ell k}\,,
\label{eq:id1}
\end{equation}
which is, therefore, a direct consequence of the
completeness relation for $\bbR$.  Note that since this derivation makes no use of the explicit form of $\QQB$, its result applies generally to models beyond the linear moose.

We next discuss the product of three-point couplings involving a mixture of gauge and Nambu-Goldstone modes,
\begin{eqnarray}
\lefteqn{
  \sum_i (M_n + M_m)\tilde{G}_{31}^{nmi}
         (M_\ell + M_k)\tilde{G}_{31}^{\ell ki}
} \nonumber\\
  & & 
  = \frac{1}{4} \sum_i \sum_{j,j'}
     g_{j-1} g_j f_j 
     (\bbR_{j,n} \bbR_{j-1,m} - \bbR_{j-1,n} \bbR_{j,m} )
     \tilde{\bbR}_{j,i} 
  \nonumber\\
  & & \qquad \times
     g_{j'-1} g_{j'} f_{j'}
     (\bbR_{j',\ell} \bbR_{j'-1,k} - \bbR_{j'-1,\ell} \bbR_{j',k} )
     \tilde{\bbR}_{j',i}.
\end{eqnarray}
Applying the completeness relation Eq.(\ref{eq:completeness2}) yields 
\begin{eqnarray}
  \sum_i (M_n + M_m)\tilde{G}_{31}^{nmi} (M_\ell + M_k)\tilde{G}_{31}^{\ell ki}
  &=&
     \frac{1}{4} \sum_{j} g_{j-1}^2 g_j^2 f_j^2 
     (\bbR_{j,n} \bbR_{j-1,m} - \bbR_{j-1,n} \bbR_{j,m} )
     (\bbR_{j,\ell} \bbR_{j-1,k} - \bbR_{j-1,\ell} \bbR_{j,k} ).\nonumber\\
     & &
\label{eq:t31sum}
\end{eqnarray}
Furthermore, combining Eq.(\ref{eq:eigenmass-rel}) and the completeness relation
Eq.(\ref{eq:completeness1}) yields,
\begin{eqnarray}
  \sum_i G_{3}^{nki} M_i^2 G_{3}^{m\ell i}
  &=&
     \frac{1}{4} \sum_{j} g_{j-1}^4 f_j^2 
     \bbR_{j-1, n} \bbR_{j-1,m} \bbR_{j-1, \ell} \bbR_{j-1, k}
  \nonumber\\
  & & + 
     \frac{1}{4} \sum_{j} g_{j}^4 f_j^2 
     \bbR_{j, n} \bbR_{j,m} \bbR_{j, \ell} \bbR_{j, k}
  \nonumber\\
  & & - 
     \frac{1}{4} \sum_{j} g_{j-1}^2 g_{j}^2 f_j^2 
     \bbR_{j-1, n} \bbR_{j,m} \bbR_{j, \ell} \bbR_{j-1, k}
  \nonumber\\
  & & - 
     \frac{1}{4} \sum_{j} g_{j-1}^2 g_{j}^2 f_j^2  
     \bbR_{j, n} \bbR_{j-1,m} \bbR_{j-1, \ell} \bbR_{j, k}
  \,,
  \label{eq:completeness3}
\end{eqnarray}
from which one may derive,
\begin{eqnarray}
  \sum_i (G_3^{nki} G_3^{m\ell i} 
  -    G_3^{n\ell i} G_3^{mk i})M_i^2 
  &=&  \sum_i G_3^{nki} M_i^2 G_3^{m\ell i}  - (n\leftrightarrow m)\,.
\end{eqnarray}
Comparing this term by term with Eq. (\ref{eq:t31sum}) 
reveals another identity,
\begin{equation}
  \sum_i (M_n + M_m)\tilde{G}_{31}^{nmi}
         (M_\ell + M_k)\tilde{G}_{31}^{\ell ki}
  =   
  \sum_i (G_3^{nki} G_3^{m\ell i} - G_3^{n\ell i} G_3^{mk i})M_i^2 ,
\label{eq:id2}
\end{equation}
which may be regarded as a consequence of the completeness relation
for $\tilde{\bbR}$.

Note also that, thanks to  Bose symmetry,
the gauge-boson coupling $G_4^{nm\ell k}$ 
is symmetric under any exchange of its indicies 
(such as $n\leftrightarrow m$), as is $G_3^{nmk}$.
We are thus able to re-express the completeness relations for
the gauge bosons and Nambu-Goldstone bosons, Eq.(\ref{eq:id1})
and Eq.(\ref{eq:id2}), in various equivalent forms. 

\subsection{Ward-Takahashi identities}

The Nambu-Goldstone boson wave-function is related with the massive
gauge boson wave-function through Eq.(\ref{eq:WT1}).
This relation is essential for the cancellation of unphysical poles in
the physical particle scattering amplitudes; 
identities derived from Eq.(\ref{eq:WT1}) may thus be regarded as Ward-Takahashi
identities. In this section, we derive several identities among coupling constants
$G$ and $\tilde{G}$ from Eq.(\ref{eq:WT1}) in linear moose models.  The general method
used here may also be applied to more general deconstructed models, and the detailed
form of the relations obtained may depend on the form that Eq. (\ref{eq:WT1}) takes in those models.

We first consider a possible relation between $G_3^{nmk}$ and
$\tilde{G}_{31}^{nmk}$.  
Combining Eq.(\ref{eq:WT1}) with Eq.(\ref{eq:deft31}) we obtain
\begin{eqnarray}
  (M_n + M_m) \tilde{G}_{31}^{nmk} M_k 
  &=& \frac{1}{4} \sum_j g_{j-1}^2 g_j  f_j^2 
      \bbR_{j,n} \bbR_{j-1,m} \bbR_{j-1,k}
  \nonumber\\
  & &-\frac{1}{4} \sum_j g_{j-1} g_j^2  f_j^2 
      \bbR_{j,n} \bbR_{j-1,m} \bbR_{j,k}      
  \nonumber\\
  & &-\frac{1}{4} \sum_j g_{j-1}^2 g_j  f_j^2 
      \bbR_{j-1,n} \bbR_{j,m} \bbR_{j-1,k}      
  \nonumber\\
  & &+\frac{1}{4} \sum_j g_{j-1} g_j^2  f_j^2 
      \bbR_{j-1,n} \bbR_{j,m} \bbR_{j,k}    .
\label{eq:calct31}
\end{eqnarray}
Here we see $\tilde{G}_{31}^{nmk}$ is  expressed solely in terms
of $\bbR$, without using $\tilde{\bbR}$.
On the other hand, using Eq.(\ref{eq:eigenmass-rel}) it is straightforward to obtain
\begin{eqnarray}
  G_3^{nmk} M_n^2  
  &=&  \frac{1}{4} \sum_j g_j^3 f_j^2
       \bbR_{j,n}\bbR_{j,m}\bbR_{j,k}
  \nonumber\\
  & & +\frac{1}{4} \sum_j g_{j-1}^3 f_j^2
       \bbR_{j-1,n}\bbR_{j-1,m}\bbR_{j-1,k}
  \nonumber\\
  & & -\frac{1}{4} \sum_j g_{j-1}^2 g_{j} f_{j}^2 
       \bbR_{j,n}\bbR_{j-1,m}\bbR_{j-1,k} 
  \nonumber\\
  & & -\frac{1}{4} \sum_j g_{j-1} g_j^2 f_j^2 
       \bbR_{j-1,n}\bbR_{j,m}\bbR_{j,k} ,
\label{eq:calc3m}
\end{eqnarray}
from which one can obtain
\begin{equation}
  G_3^{nmk} (M_n^2  - M_m^2)
  =  G_3^{nmk} M_n^2 - (n\leftrightarrow m)\,.
\label{eq:calc3}
\end{equation}
Comparing Eq.(\ref{eq:calc3}) with Eq.(\ref{eq:calct31}), we find,
\begin{equation}
  (M_n + M_m) \tilde{G}_{31}^{nmk} M_k 
  = - G_3^{nmk} (M_n^2 - M_m^2),
\label{eq:WT-1}
\end{equation}
which is the desired relation between $\tilde{G}_{31}$ and $G_3$,

We next study a relation between $\tilde{G}_3$ and $G_3$.
Using Eq.(\ref{eq:WT1}) twice, we find
\begin{eqnarray}
  \tilde{G}_3^{nmk} M_n M_m 
  &=& \frac{1}{4}\sum_j g_{j-1}^3 f_j^2 
      \bbR_{j-1,n} \bbR_{j-1,m} \bbR_{j-1,k}
     +\frac{1}{4}\sum_j g_{j-1}^2 g_j f_j^2 
      \bbR_{j-1,n} \bbR_{j-1,m} \bbR_{j,k}
  \nonumber\\
  & & +\frac{1}{4} \sum_j g_{j}^3 f_j^2 
      \bbR_{j,n} \bbR_{j,m} \bbR_{j,k}
     +\frac{1}{4}\sum_j g_{j-1} g_j^2 f_j^2 
      \bbR_{j,n} \bbR_{j,m} \bbR_{j-1,k}
  \nonumber\\
  & &-\frac{1}{4} \sum_j g_{j-1} g_{j}^2 f_j^2 
      \bbR_{j-1,n} \bbR_{j,m} \bbR_{j,k}
     -\frac{1}{4}\sum_j g_{j-1}^2 g_j f_j^2 
      \bbR_{j-1,n} \bbR_{j,m} \bbR_{j-1,k}
  \nonumber\\
  & &-\frac{1}{4} \sum_j g_{j-1} g_{j}^2 f_j^2 
      \bbR_{j,n} \bbR_{j-1,m} \bbR_{j,k}
     -\frac{1}{4}\sum_j g_{j-1}^2 g_j f_j^2 
      \bbR_{j,n} \bbR_{j-1,m} \bbR_{j-1,k}.
  \nonumber\\
\label{eq:calct3a}
\end{eqnarray}
Using Eq.(\ref{eq:calc3m}), it is straightforward to calculate
\begin{equation}
  G_3^{nmk}(M_n^2 + M_m^2 - M_k^2) = G_3^{nmk} M_n^2 +
  (n\leftrightarrow m) - (n \leftrightarrow k).
\label{eq:calc3b}
\end{equation}
Comparing the result of Eq.(\ref{eq:calc3b}) with
Eq.(\ref{eq:calct3a}) yields
an identity between $\tilde{G}_3$ and $G_3$,
\begin{equation}
  \tilde{G}_3^{nmk} M_n M_m = G_3^{nmk}(M_n^2 + M_m^2 - M_k^2).
\label{eq:WT-2}
\end{equation}

We next consider identities which involve the quartic $\pi$ coupling
$\tilde{G}_4$.
Using Eq.(\ref{eq:WT1}) we see
\begin{equation}
  \tilde{G}_4^{nm\ell k} M_k 
  = \frac{1}{2} \sum_j \dfrac{v^2}{f_j} 
    \tilde{\bbR}_{j,n}\tilde{\bbR}_{j,m}\tilde{\bbR}_{j,\ell}
    (g_{j-1} \bbR_{j-1,k}- g_j \bbR_{j,k}).
\end{equation}
Comparing this with the definition of $\tilde{G}_{41}$
Eq.(\ref{eq:deft41}), we immediately find the identity
\begin{equation}
  \tilde{G}_4^{nm\ell k} M_k = - \frac{v}{2} \tilde{G}_{41}^{nm\ell k} . 
\label{eq:WT-3}
\end{equation}
which is  recognizable as a generalization of Eq. (\ref{eq:G41a}).

It is also straightforward to start from an expression involving  two powers of mass  and the four-point Nambu-Goldstone coupling
\begin{equation}
  \tilde{G}_4^{nm\ell k} M_\ell M_k 
  = \frac{v^2}{4} \sum_j 
    \tilde{\bbR}_{j,n}\tilde{\bbR}_{j,m}
    (g_{j-1} \bbR_{j-1,\ell}- g_j \bbR_{j,\ell})
    (g_{j-1} \bbR_{j-1,k}- g_j \bbR_{j,k}).
\label{eq:tG4MM}
\end{equation}
Applying the completeness relation for $\bbR$,
Eq.(\ref{eq:completeness1}), gives
\begin{equation}
  \sum_i \tilde{G}_3^{nmi} G_3^{\ell ki} 
  = \sum_j \tilde{\bbR}_{j,n} \tilde{\bbR}_{j,m}
    (g_j^2 \bbR_{j,\ell} \bbR_{j,k} 
   + g_{j-1}^2 \bbR_{j-1,\ell} \bbR_{j-1,k}).
\label{eq:completeness4}
\end{equation}
Comparing Eq.(\ref{eq:tG4MM}), Eq.(\ref{eq:completeness4}), and the
definition of $\tilde{G}_{42}$ Eq.(\ref{eq:deft42}),
we obtain an identity
\begin{equation}
  \tilde{G}_4^{nm\ell k} M_\ell M_k 
  = \dfrac{v^2}{2} \tilde{G}_{42}^{nm\ell k}
   + \dfrac{v^2}{4} \sum_i \tilde{G}_3^{nmi} G_3^{\ell ki} .
\label{eq:WT-4}
\end{equation}
that relates $\tilde{G}_4$ to $\tilde{G}_{42}$ and the triple-gauge coupling.

Starting from an expression involving four powers of mass  and the four-point Nambu-Goldstone coupling produces yet another identity.  Using Eq.(\ref{eq:WT1}) four times, we obtain
\begin{eqnarray}
  \tilde{G}_4^{nm\ell k} M_n M_m M_\ell M_k 
  &=& 
  \dfrac{v^2}{16} \sum_j f_j^2 
    (g_{j-1} \bbR_{j-1,n} - g_j \bbR_{j,n})
    (g_{j-1} \bbR_{j-1,m} - g_j \bbR_{j,m}) \times
  \nonumber\\
  & & \qquad \quad \times
    (g_{j-1} \bbR_{j-1,\ell} - g_j \bbR_{j,\ell})
    (g_{j-1} \bbR_{j-1,k} - g_j \bbR_{j,k}).
\end{eqnarray}
After some lengthy calculation using Eq.(\ref{eq:eigenmass-rel}) and
Eq.(\ref{eq:completeness3}), we obtain 
\begin{eqnarray}
  \tilde{G}_{4}^{nm\ell k} M_n M_m M_\ell M_k  
  &=& \dfrac{v^2}{4} G_4^{nm\ell k}(M_n^2 + M_m^2 + M_\ell^2 + M_k^2)
  \nonumber\\
  & & -\dfrac{v^2}{4} \sum_j (G_3^{nmj}G_3^{\ell kj} + G_3^{n\ell j} G_3^{mkj} 
          +G_3^{nkj}G_3^{m\ell j} ) M_j^2 .
\label{eq:WT-5}
\end{eqnarray}

The set of  identities we have just derived, Eq.(\ref{eq:WT-1}),
Eq.(\ref{eq:WT-2}), Eq.(\ref{eq:WT-3}), Eq.(\ref{eq:WT-4}) and Eq.(\ref{eq:WT-5}),
are interesting
because they completely determine the Nambu-Goldstone couplings
($\tilde{G}_{31}$, $\tilde{G}_3$, $\tilde{G}_{41}$, $\tilde{G}_{42}$
and 
$\tilde{G}_4$) solely in terms of the gauge-boson couplings.  We now rewrite these deconstruction  identities in a form that emphasizes this property (see also Table \ref{tab:wtcr}):
\begin{eqnarray}
  (M_n + M_m)\tilde{G}_{31}^{nmk} M_k 
  &=& - G_3^{nmk} (M_n^2 - M_m^2), 
\label{eq:WT-t31}
  \\ 
  \tilde{G}_{3}^{nmk} M_n M_m 
  &=& G_3^{nmk}(M_n^2 + M_m^2 - M_k^2),
\label{eq:WT-t3}
  \\
  \tilde{G}_{42}^{nm\ell k} M_n M_m
  &=& \frac{1}{2} G_4^{nm\ell k}(M_\ell^2 + M_k^2)
     -\frac{1}{2} \sum_i 
     (G_3^{n\ell i} G_3^{mki} + G_3^{nki}G_3^{m\ell i})M_i^2,
  \nonumber\\
  & &
\label{eq:WT-t42}
  \\
  \tilde{G}_{41}^{nm\ell k} M_n M_m M_\ell
  &=&-\frac{v}{2} G_4^{nm\ell k}(M_n^2 + M_m^2 + M_\ell^2 + M_k^2)
  \nonumber\\
  & &
     +\frac{v}{2} \sum_i 
      (G_3^{nmi} G_3^{\ell k i} 
      +G_3^{n\ell i} G_3^{m k i}
      +G_3^{nki} G_3^{m \ell i}) M_i^2,
\label{eq:WT-t41}
  \\
  \tilde{G}_{4}^{nm\ell k} M_n M_m M_\ell M_k
  &=&
     \dfrac{v^2}{4} 
     G_4^{nm\ell k}(M_n^2 + M_m^2 + M_\ell^2 + M_k^2)
  \nonumber\\
  & &
     -\dfrac{v^2}{4} \sum_i 
      (G_3^{nmi} G_3^{\ell k i} 
      +G_3^{n\ell i} G_3^{m k i}
      +G_3^{nki} G_3^{m \ell i}) M_i^2.
\label{eq:WT-t4}
\end{eqnarray}
The minor manipulations involved in reaching the revised form of the deconstruction identities is readily apparent in most cases.  However in the case of  Eq.(\ref{eq:WT-t42}), several distinct steps are involved.  We started from Eq. (\ref{eq:WT-4}) and multiplied it  by $M_n M_m$ so as to be able to eliminate $\tilde{G_4}$ using Eq. (\ref{eq:WT-t4}).  Next, we used Eq. (\ref{eq:WT-t3}) to eliminate $\tilde{G}_3$.  Finally, we invoked the gauge-boson completeness relation, Eq.(\ref{eq:id1}), to simplify the result.

\subsection{Comparison with results from the Equivalence Theorem}

We now show that the generalized ET sum rules can be understood by using 
Ward-Takahashi identities  Eqs.(\ref{eq:WT-t31})--(\ref{eq:WT-t4}) 
and the completeness relation sum rules
Eqs.(\ref{eq:id1})--(\ref{eq:id2}).  We will run through the ET sum rules
in the order in which they appear in Table \ref{tab:sumrules}.

We start with the pair of ET sum rules, Eqs. (\ref{eq:E4-sum1}) and (\ref{eq:E4-sum2}), relating the triple and quartic gauge couplings. 
\begin{equation}
  \sum_i G_3^{nni} G_3^{mmi} = G_4^{nnmm}\,, \qquad\qquad
  \sum_i G_3^{nmi} G_3^{nmi} = G_4^{nnmm}. \nonumber
\end{equation}
Using Bose symmetry, $G_4^{nnmm}=G_4^{nmnm}$ and
$G_3^{nmi}=G_3^{mni}$, we find these can
be derived from the gauge-boson completeness relation for $\bbR$, Eq.(\ref{eq:id1}).

The next pair of ET sum rules in the table, Eqs. (\ref{eq:E0-sum2}) and (\ref{eq:SR-E0-LLTT-2in}),
\begin{eqnarray}
   &G_4^{nnmm}&(M_n^2-M_m^2)^2 =
 \nonumber\\
   \qquad & & \qquad \sum_i [(G_3^{nmi})^2 - G_3^{nni} G_3^{mmi}]M_i^2(2M_n^2+2M_m^2-M_i^2)
  +M_n^2 M_m^2 \sum_i [
    (\tilde{G}_3^{nmi})^2 - \tilde{G}_3^{nni}\tilde{G}_3^{mmi}],
  \nonumber\\
    &G_4^{nnmm}& M_n^2 =
  \frac{1}{2} \sum_i \left[\tilde{G}_3^{nni} G_3^{mmi} M_n^2 
  +G_3^{nni} G_3^{mmi} M_i^2 \right],
\nonumber
\end{eqnarray}
are also related to the gauge boson completeness relation for $\bbR$.  If one applies
the deconstruction identity for $\tilde{G}_3$, Eq. (\ref{eq:WT-t3}), to each of these sum rules, 
in order to eliminate the factors of $\tilde{G}_3$, one obtains, respectively,
\begin{equation}
  G_4^{nnmm}(M_n^2-M_m^2)^2
  = \sum_i (G_3^{nmi})^2 (M_n^2 + M_m^2)^2 - 4 \sum_i
  G_3^{nni}G_3^{mmi} M_n^2 M_m^2 ,
\label{eq:hjh02r}
\end{equation}
and
\begin{equation}
  G_4^{nnmm} M_n^2 = \sum_i G_3^{nni} G_3^{mmi} M_n^2.
\end{equation}
Both of these equations can be verified using the completeness relation for $\bbR$,
Eq.(\ref{eq:id1}), and the Bose symmetry of $G_4^{nnmm}$.

The next two ET sum rules listed in Table \ref{tab:sumrules}, Eqs. (\ref{eq:E2-sum1})-(\ref{eq:E2-sum2}), 
are each related to the Nambu-Goldstone-boson completeness relation for $\tilde\bbR$, Eq.(\ref{eq:id2}).  To see this, we must first use the deconstruction identity for $\tilde{G}_{31}$, Eq.(\ref{eq:WT-t31}), to eliminate the Nambu-Goldstone boson coupling from  Eq.(\ref{eq:id2}).  This yields a relation of the form,
\begin{equation}
  (M_n^2-M_m^2)(M_\ell^2-M_k^2) \sum_i G_{3}^{nmi} G_{3}^{\ell k i} 
      \dfrac{1}{M_i^2}
  =   
  \sum_i (G_3^{nki} G_3^{m\ell i} - G_3^{n\ell i} G_3^{mk i})M_i^2 \,,
\label{eq:sumrule2}
\end{equation}
which has not previously appeared in the literature.  Now we can see directly that Eq. (\ref{eq:E2-sum2}),
\begin{equation}
  \sum_i\left[( G_3^{nmi})^2-G_3^{nni}G_3^{mmi}\right] M_i^2
 =(M_n^2-M_m^2)^2\sum_i (G_3^{nmi})^2 \dfrac{1}{M_i^2}\, ,\nonumber
\end{equation}
 is a special case of this new relation, with $\ell=n$
and $k=m$.  Similarly, let us eliminate the Nambu-Goldstone coupling $\tilde{G}_4^{nnmm}$
from Eq. (\ref{eq:E2-sum1}),
\begin{equation}
  2(M_n^2+M_m^2) G_4^{nnmm} + \sum_i (G_3^{nmi})^2 \left[
    \dfrac{(M_n^2-M_m^2)^2}{M_i^2}- 3 M_i^2\right]
  = \dfrac{4M_n^2 M_m^2}{v^2} \tilde{G}_4^{nnmm}  \,,
\nonumber
\end{equation}
by applying  the deconstruction 
identity for $\tilde{G}_4$, Eq.(\ref{eq:WT-t4}).  The resulting equation,
\begin{equation}
  (M_n^2-M_m^2)^2\sum_i (G_3^{nmi})^2 \dfrac{1}{M_i^2}
 =\sum_i (G_3^{nmi}G_3^{nmi}-G_3^{nni}G_3^{mmi}) M_i^2\,,
\end{equation}
is also a special case of the Nambu-Goldstone-boson completeness relation, Eq.(\ref{eq:sumrule2}), for $\ell=n$
and $k=m$. 

The next pair  ET sum rules listed in the table, Eqs. (\ref{eq:E0-sum1}) and (\ref{eq:SR-E0-LLTT-1in}),
\begin{eqnarray}
 M_n^2 M_m^2 \sum_i (\tilde{G}_3^{nmi})^2  &=&   \sum_i (G_3^{nmi})^2 (M_n^2 + M_m^2 - M_i^2)^2 
  \nonumber\\
  \sum_i (\tilde{G}_{3}^{nim})^2 M_n^2 
 & = &\sum_i (G_{3}^{nmi})^2 \dfrac{(M_i^2 + M_n^2 - M_m^2)^2}{M_i^2},
\nonumber
\end{eqnarray}
are each satisfied term by term for every
KK level $i$ in the sum, due to the deconstruction identity for $\tilde{G}_3$, Eq.(\ref{eq:WT-t3}).

The next equality obtained from the ET, Eq. (\ref{eq:G41a}),
\begin{equation}
  \tilde{G}_{41}^{nnmm} = - \dfrac{2M_m}{v} \tilde{G}_4^{nnmm}
\nonumber
\end{equation}
follows directly from the pair of deconstruction identities for $\tilde{G}_{41}$ and $\tilde{G}_4$,
Eq.(\ref{eq:WT-t41}) and Eq.(\ref{eq:WT-t4}).

The final ET sum rule in the table, Eq.(\ref {eq:SR-E0-LLTT-3in}),
\begin{equation}
  \tilde{G}_{42}^{nnmm} M_n^2 = G_4^{nnmm} M_m^2 
  - \sum_i (G_3^{nmi})^2 M_i^2\,,
\nonumber
\end{equation}
 can be proven by using the deconstructionidentity for $\tilde{G}_{42}$, Eq.(\ref{eq:WT-t42}). 
 
 We have now seen that all of the ET sum rules can be rederived by using the
 Ward-Takahashi identities and completeness relations.

\section{ Conclusions}
\label{sec:concl}

\subsection{Summary}

We have derived a general set of sum rules among the gauge boson couplings, NG-boson couplings and KK gauge boson masses in Higgsless models of electroweak symmetry breaking.  As shown in Table \ref{tab:sumrules}, these sum rules ensure the cancellation of bad high-energy behavior in elastic and inelastic $nn \to mm$ scattering processes involving gauge and NG bosons.  As summarized in Table \ref{tab:wtcr}, even more general forms of these relations may be formulated as a combination of completeness relations for the gauge and NG boson KK wavefunctions and Ward-Takashi identities relating NG boson couplings to pure gauge couplings.

Of the many sum rules and identities we have derived, the completeness relations 
for the gauge bosons and Nambu-Goldstone bosons, written in terms of the gauge
boson couplings \footnote{The NG boson completeness relation (\ref{eq:id2}) may be so written by using the Ward-Takashi identity Eq. (\ref{eq:WT-t3}) to eliminate $\tilde{G_3}$.} as in 
Eqs.(\ref{eq:id1}) and (\ref{eq:sumrule2}), are particularly important, since they are
relevant for the scattering amplitudes of physically polarized gauge-bosons.
In fact, Eq.(\ref{eq:id1}) is an inelastic generalization of the
first sum rule for elastic scattering 
\begin{equation}
  G_4^{nnnn} = \sum_i G_4^{nni} G_4^{nni}
\label{eq:elastic1}
\end{equation}
that was discussed in Ref.\cite{Csaki:2003dt} as ensuring the cancellation of the
${\cal O}(E^4)$ behavior in the elastic $W_LW_L$ scattering amplitude. We have
seen that the generalized form additionally ensures cancellation of ${\cal O}(E^3)$ terms in
$W_L W_L \to W_L W_T$ scattering, of ${\cal O}(E^2)$ terms in
$W_L W_L \to W_T W_T$ scattering, and ${\cal O}(E^1)$ terms in
$W_L W_T \to W_T W_T$ processes.  In contrast, the sum rule, Eq.(\ref{eq:sumrule2}), or its parent completeness relation (\ref{eq:id2}), applies
only for inelastic scattering because it vanishes when the KK levels of the incoming ($n$) and outgoing ($m$) states are the same.  This is a new result, not previously discussed in the literature.

Another of particular interest is the Ward-Takahashi identity for the quartic NG-boson coupling $\tilde{G}_4$.  In order to prevent bad ${\cal O}(E^2)$ high energy behavior in
the $W_L W_L$ scattering amplitudes, this coupling must vanish in the continuum limit.  Combining these pieces of information leads to a new sum rule for the four-point and three-point gauge boson couplings:
\begin{equation}
  G_4^{nm\ell k}(M_n^2 + M_m^2 + M_\ell^2 + M_k^2)
 = \sum_i 
      (G_3^{nmi} G_3^{\ell k i} 
      +G_3^{n\ell i} G_3^{m k i}
      +G_3^{nki} G_3^{m \ell i}) M_i^2,
\end{equation}
which can be regarded as an inelastic generalization of the second
sum rule for elastic scattering
\begin{equation}
  4 G_4^{nnnn} M_n^2  = 3 \sum_i G_3^{nni} G_3^{nni} M_i^2
\label{eq:elastic2}
\end{equation}
discussed in Ref.\cite{Csaki:2003dt}.

So far we have considered the case of an $SU(2)^{N+1}$ deconstructed Higgsless model,
as illustrated in Fig. \ref{fig:Fig2}. Such a model, however, corresponds to an electroweak sector
in the absence of hypercharge -- a theory with degenerate $W$ and $Z$ bosons and no photon.
A realistic deconstructed model would be based on an $SU(2)^{N+1} \times U(1)^M$ gauge
symmetry \cite{SekharChivukula:2004mu} (with $M\ge 1$). 
In such a model, the charged- and neutral-boson
mass matrices would be different, with the former being $(N+1) \times (N+1)$ dimensional
and having no massless modes and the latter $(N+M+1) \times (N+M+1)$ dimensional
and including the massless photon. Let us label the charged bosons $W^\pm_n$ ($n=0,\ldots,N$),
as above, and the neutral bosons $Z_{k'}$ where $k'=\gamma, 0, \ldots, N+M-1$.
Note that the vertices which contribute to gauge-boson
scattering in unitary gauge arise solely from the non-Abelian couplings of the $SU(2)^{N+1}$
gauge interactions of the theory, and therefore are of the form $W^+_n W^-_m Z_{k'}$, 
$W^+_nW^-_m W^+_p W^-_q$, or $W^+_nW^-_m Z_{k'} Z_{l'}$. These are precisely of the
same form that we considered in our analysis above, with the modification that
there are more neutral-bosons than charged-bosons. Consider first the 
scattering amplitude for  $W^+_n W^-_n \to Z_{k'} Z_{k'}$. This amplitude only
receives contributions from $t$- and $u$-channel $W_m$ exchange, and from the
four-point $W^+_nW^-_n Z_{k'} Z_{k'}$ coupling. The form of the amplitudes, therefore,
is precisely the same as those derived above and the relevant sum rules in Table \ref{tab:sumrules}
continue to apply as written, with all sums running over the charged boson
states. By contrast, the amplitudes $W^+_n W^-_n \to W^+_m W^-_m$ receive contributions
from $s$- and $t$-channel $Z_{k'}$ exchange (including the photon), and therefore
the relevant sum rules in Table \ref{tab:sumrules} apply with the sums running over the
neutral bosons instead. Similarly, the only non-trivial Nambu-Goldstone boson interactions
will arise from the $SU(2)^{N+1}$ non-lineary sigma-model links of the theory, but the rotation
matrix describing which link fields are eaten by which mass-eigenstate fields will differ
between the charged- and neutral-boson sectors. Again, in calculating the scattering
amplitudes one will arrive at the sum rules summarized in Table \ref{tab:sumrules} -- however,
one must be careful to interpret the sums that appear according to whether charged-
or neutral-bosons appear as intermediate states. The application of the sum rules
to an $SU(2)^{N+1} \times U(1)^M$ deconstructed Higgsless theory, therefore, is straightforward.
In a future publication \cite{Newpaper}, 
we will discuss an $SU(2)^2 \times U(1)^2$ model and illustrate how these
results are realized in a simple case.

\subsection{Generalizations}

As mentioned earlier, the sum rules derived in Section IV and listed in Table \ref{tab:sumrules} are general relations between the multi-boson couplings, viewed simply as coefficients in the effective Lagrangian and without reference to the form of $\CD_W$.  In contrast, many of the results of Section V depend on the form of $\CD_W$ because they draws directly upon the form of the mixed gauge/Nambu-Goldstone couplings in Eqs. (\ref{eq:deft41}) - (\ref{eq:deft3}).  That is, with the exception of the gauge-boson completeness relation, the identities in Table \ref{tab:wtcr} are specific to linear moose models and their 5D continuum limits.     Here, we briefly mention how the analysis of Section V could be modified to describe more general classes of models.

The simplest generalization of the linear moose involves adding one link between the two boundary groups, closing the moose into a ring.  This does not change the form of any of the couplings, but makes the sums over links periodic (e.g. in the Nambu-Goldstone kinetic term and the definitions of the multi-boson couplings involving Nambu-Goldstone modes).  As a result, the form of all of the identities in Table \ref{tab:wtcr} would be unchanged, but any sums would now be periodic.

More broadly, one could consider models in which one gauge site might be touched by more than two links or in which non-linear sigma models link pairs of gauge sites that are not  nearest-neighbors; this would include planar meshes or even non-planar configurations that could be constructed out of ``toobers and zots".  In these models, the form of the covariant derivative for the link fields must be modified to reflect the broader array of possibilities; as a result, the matrix $\CD_W$ appearing inside the matrix $\QQB$ will have additional non-zero entries.  The $\QQB$ matrix itself will no longer be square if the number of links exceeds the number of sites, meaning that the rank of $\MMWT$ will exceed that of $\MMW$; in other words, some massless Nambu-Goldstone modes (corresponding to closed loops of links) will remain in addition to those ``eaten" by the gauge fields.   The modification of the covariant derivative will alter the form of the Nambu-Goldstone couplings $\tilde{G}_3, \, \tilde{G}_{31},\, \tilde{G}_{41},$ and $\tilde{G}_{42}$; the multi-gauge couplings $G_3$ and $G_4$ and the four-Nambu-Goldstone-mode coupling $\tilde{G}_4$ will remain unchanged because they are independent of $\CD_W$.  The upshot is that there will still be relationships (Ward-Takahashi identities) between the gauge and Nambu-Goldstone couplings, but the precise forms of the relationships (e.g. the appearance of particular factors of mass or the need to sum over certain indices) will generally be altered.

To make this more concrete, let us introduce a slightly modified notation in which we
distinguish between the indices denoting sites ($a$) and those denoting links ($\hat{a}$), and likewise between the indices labeling gauge ($\alpha$) and Nambu-Goldstone ($\hat{\alpha}$) KK modes.  Thus, the matrix $\RRB_{a,\alpha}$ carries indices of site and gauge KK mode, while 
$\RRBT_{\hat{a},\hat{\alpha}}$ carries indices of link and Nambu-Goldstone KK mode. The mass matrix $\MM_{\hat{\alpha},\alpha}$ connects the massive Nambu-Goldstone and gauge KK modes, while 
$\QQB_{\hat{a},a}$ shows which links attach to which sites. 

In this notation, one would write the four-point Nambu-Goldstone coupling as
\beq
\tilde{G}_4^{\hat{\alpha},\hat{\beta},\hat{\gamma},\hat{\delta}} = \sum_{\hat{a}} \frac{v^2}{f^2_{\hat{a}}} \tilde{R}_{\hat{a},\hat{\alpha}}\,\tilde{R}_{\hat{a},\hat{\beta}}\, \tilde{R}_{\hat{a},\hat{\gamma}}\, \tilde{R}_{\hat{a},\hat{\delta}}
\eeq
This is a coupling that does not depend on the form of $\CD_W$ or $\QQB$.  Likewise, the relationship between $\tilde{G}_4^{nnmm}$ and $\tilde{G}_{41}^{nnmm}$ was derived from the Equivalence Theorem and does not depend on the form of $\CD_W$.  Accordingly, we have:
\beq
\tilde{G}_4^{\hat{\alpha},\hat{\alpha},\hat{\gamma},\hat{\gamma}} \MMB_{\hat{\gamma},\gamma} = -\frac{v}{2} \tilde{G}_{41}^{\hat{\alpha},\hat{\alpha},\hat{\gamma},\hat{\gamma}} 
\eeq
However, the specific form of $\tilde{G}_{41}$ does depend on $\QQB$ because Eq. (\ref{eq:rtmqr}) is used to relate $\tilde{G}_4$ to $\tilde{G}_{41}$:
\beq
\tilde{G}_4^{\hat{\alpha},\hat{\alpha},\hat{\gamma},\hat{\gamma}} = \sum_{\hat{a}} \frac{v^2}{f^2_{\hat{a}}} \tilde{R}_{\hat{a},\hat{\alpha}}\,\tilde{R}_{\hat{a},\hat{\alpha}}\, \tilde{R}_{\hat{a},\hat{\gamma}}\, \tilde{R}_{\hat{a},\hat{\gamma}}\MMB_{\hat{\gamma},\gamma} =  \sum_{\hat{a}} \frac{v^2}{f^2_{\hat{a}}} \tilde{R}_{\hat{a},\hat{\alpha}}\,\tilde{R}_{\hat{a},\hat{\alpha}}\, \tilde{R}_{\hat{a},\hat{\gamma}}\, 
Q_{\hat{a},a} R_{a, \gamma} = -\frac{v}{2} \tilde{G}_{41}^{\hat{\alpha},\hat{\alpha},\hat{\gamma},\hat{\gamma}} 
\eeq
The form of $\tilde{G}_{41}$ given for the linear deconstructed moose in Eq. (\ref{eq:deft41}) includes a factor of $(g_j \bbR_{j,k} - g_{j-1} \bbR_{j-1,k})$\,, which corresponds to the product $\QQB\RRB$ above (modulo a factor of $\frac{v}{f_{\hat{a}}}$). Generally speaking, we can see that  when a multi-Nambu-Goldstone vertex is acted on by a factor of $M$, the resulting product $\MMB\RRBT$ becomes $\QQB\RRB$, changing a Nambu-Goldstone index into a gauge index.  Hence, the relationship between $\tilde{G}_4$ and $\tilde{G}_{42}$ in Eq. (\ref{eq:WT-4}) involves two factors of $\QQB\RRB$. 

\subsection{Consistency of the Continuum Limit}

 Ref.\,\cite{Birkedal:2004au} considered
 $WZ$ scattering in a 5D continuum Higgsless scenario by assuming that the contributions of
 all higher KK modes except $(W_1,\,Z_1)$ could be ignored.  The general sum rules we have derived in this paper demonstrate that such  assumption is generally {\it inconsistent} in 5D once scattering in
 inelastic channels becomes important.

 We start by considering the elastic scattering of zero-modes:
 $W_{0L}^aW_{0L}^b\to W_{0L}^cW_{0L}^d$
 ($\pit_0^a\pit_0^b\to \pit_0^c\pit_0^d$). From the elastic sum
 rules (\ref{eq:E4-sum-el}), we have an expression for the quartic gauge coupling in terms of triple-gauge couplings
 \beqa
 \label{eq:E4sum-el-N+1=2}
 G_4^{0000} &=& \sum_{k} \(G_3^{00k}\)^2\,.
 \eeqa
 On the other hand, the sum rule in Eq. (\ref{eq:E2-sum-el})
 gives a different expression for the quartic gauge coupling
 \beqa
 \label{eq:E2sum-el-N+1=2}
 G_4^{0000}
 &=& \f{3}{4}\[\sum_{k} \(G_3^{00k}\)^2\f{M_k^2}{M_0^2}\] +
     \f{M_0^4}{v^2}\Gt_4^{0000}  \,.
 \eeqa
 Combining (\ref{eq:E4sum-el-N+1=2}) and (\ref{eq:E2sum-el-N+1=2}) provides
 an expression for the triple gauge boson coupling among three zero-modes
 \beqa
 \label{eq:G3-000}
 \(G_3^{000}\)^2 &=&
 \(G_3^{001}\)^2 \(\frac34 \frac{M^2_1}{M_0^2} - 1\) + 
 \sum_{n \geq 2}  \(G_3^{00n}\)^2 \(\frac34 \frac{M^2_n}{M_0^2} - 1\) 
 + \frac{M_0^4}{v^2}\tilde{G}_4^{0000}
 \eeqa
 which can still be satisfied even if the quartic Nambu-Goldstone coupling is set to zero
as in the continuum limit.\,Here we have separated out the higher KK modes ($n \geq 2$) for later convenience.
 
Now, let us consider the inelastic channel
 $W_{0L}^aW_{0L}^b\to W_{1L}^cW_{1L}^d$
 ($\pit_0^a\pit_0^b\to \pit_1^c\pit_1^d$).
 This time, the inelastic sum
 rules of Eqs. (\ref{eq:E4-sum1})-(\ref{eq:E4-sum2}) give one expression for $G_4^{0011}$
 \beqa
 \label{eq:E4sum-inel-N+1=2}
 G_4^{0011} &=&  \sum_{k} \(G_3^{01k}\)^2\,,
 \eeqa
 while the inelastic sum rule (\ref{eq:E2-sum1}) yields another:
 \beqa
 \label{eq:E2sum-inel-N+1=2}
 2 (M_0^2 + M_1^2) G_4^{0011} = - \sum_{k}\(G_3^{01k}\)^2
 \[\f{(M_1^2 - M_0^2)^2}{M_k^2} - 3M_k^2\] + \frac{4 M_0^2 M_1^2}{v^2}\tilde{G}_4^{0011}\,.
 \eeqa
 Combining Eqs. (\ref{eq:E4sum-inel-N+1=2}) and (\ref{eq:E2sum-inel-N+1=2}),
 we may solve for $\tilde{G}_4^{0011}$,
 \beqa
 \label{eq:Gt4-0011}
 \frac{4 M_0^2 M_1^2}{v^2} \tilde{G}_4^{0011} &=& \(G_3^{010}\)^2 \frac{M_1^4}{M_0^2}
 + \(G_3^{011}\)^2 \frac{M_0^4}{M_1^2}
 + \sum_{n\geq 2} \(G_3^{01n}\)^2 \[ -3 M_n^2 + 2\(M_0^2 +M_1^2\) + \frac{\(M_1^2 - M_0^2\)^2}{M_n^2} \]\,,
 \eeqa
 where we have, again, separated out the higher KK modes ($n \geq 2$) for our convenience.

At this point, an inconsistency emerges if we simultaneously take the continuum limit (by setting $\tilde{G}_4 = 0$) and include only the modes with KK number $n = 0,1$ in the inelastic scattering process.  Specifically, setting $\tilde{G}_4^{0011} = 0$ and dropping the terms for $n \geq 2$ in Eq. (\ref{eq:Gt4-0011}) leads to an equality
\beqa
0 &=&  \(G_3^{010}\)^2 \frac{M_1^4}{M_0^2}
 + \(G_3^{011}\)^2 \frac{M_0^4}{M_1^2}
\eeqa 
that can only be satisfied if $G_3^{010} = G_3^{011} = 0$; due to Bose symmetry, $G_3^{001}$ must also equal zero.   But this scenario causes a problem in our equation for the elastic process above, Eq. (\ref{eq:G3-000}); if we simultaneously set $G_3^{001} = 0$, set $\tilde{G}_4^{0000} = 0$ and drop the terms for $n \geq 2$, the triple-gauge coupling $G^{000}_3$ for the zero modes (i.e. the WWZ coupling) must vanish.

As the full form of equations (\ref{eq:G3-000}) and (\ref{eq:Gt4-0011}) suggests, this inconsistency can be avoided by retaining the higher KK modes when taking the continuum limit. See \cite{He:2007ge} for a discussion of some phenomenological implications.

\vspace{.5cm}
\centerline{\bf Acknowledgments}

R.S.C. and E.H.S. are supported in part by the US NSF under
grant  PHY-0354226 and also acknowledge support from the CERN Theory Institute.  M.T.'s work is supported in part by the JSPS Grant-in-Aid for Scientific Research No.20540263. H.J.H. is supported by the NSF of China under grants 10625522 and 10635030.  M.K. is supported in part by the US NSF under grant PHY-00-98527.  The authors acknowledge the support of the Radcliffe Institute for Advanced Study, the Kavli Institute for Theoretical Physics (Santa Barbara), the Kavli Institute for Theoretical Physics - China (Beijing), and the Aspen Center for Physics during the completion of this work.

 \appendix
 \section{Application to the Three-Site Model}

In this appendix, we apply our results to the 
 simplest deconstructed Higgsless model \cite{SekharChivukula:2006cg}, which incorporates only three
sites on the deconstructed lattice.  The only additional vector states 
(other than the usual electroweak gauge bosons) are a 
triplet of $W_1$ and $Z_1$ bosons, which
may be interpreted as the lightest Kaluza-Klein states of a compactified five-dimensional
theory. This theory is in the same class as models of extended  electroweak gauge symmetries \cite{Casalbuoni:1985kq,Casalbuoni:1996qt}  motivated by models of hidden local symmetry \cite{Bando:1985ej,Bando:1985rf,Bando:1988ym,Bando:1988br,Harada:2003jx} in QCD, and the gauge sector is precisely that of the BESS model \cite{Casalbuoni:1985kq}. While simple, the three site model \cite{SekharChivukula:2006cg} is
sufficiently rich to describe the physics associated with fermion mass generation,
as well as the  fermion delocalization \cite{Cacciapaglia:2004rb,Cacciapaglia:2005pa,Foadi:2004ps,Foadi:2005hz,Chivukula:2005bn,Casalbuoni:2005rs,SekharChivukula:2005xm}
required in order to accord with precision electroweak tests
\cite{Peskin:1992sw,Altarelli:1990zd,Altarelli:1991fk,Barbieri:2004qk,Chivukula:2004af}.  Detailed investigations of the one-loop chiral logarithmic corrections to the $S$ and $T$ parameters have been reported in     \cite{Matsuzaki:2006wn,Chivukula:2007ff,Dawson:2007yk}.

 \subsection{ Masses}

 In this subsection we will first give the gauge boson mass-matrix and its dual; and
 we then diagonalize them and derive the relevant  quartic and triple boson 
 couplings. For the purpose of demonstrating the sum rules, we will still
 set \,$g_2^2\simeq {g'}^2\simeq 0$\, in this model.
 Thus we have only two gauge couplings for the analysis
 $\,(\gz ,\,g_1)$; we define the ratio $\,x\equiv g_0/g_1\,$ for later convenience, as in \cite{SekharChivukula:2006cg}. The 
 two moose links are associated with the Nambu-Goldstone boson decay
 constants $(f_1,\,f_2)$, and we also define the ratio
 $\,y\equiv f_2/f_1\,$ for convenience. The symmetry breaking
 pattern leads to the relations\,\cite{SekharChivukula:2006cg}
 \begin{equation}
 \label{eq:gv}
 \f{1}{g_0^2} + \f{1}{g_1^2} ~=~ \f{1}{g^2}\,, \qquad\qquad 
 \f{1}{f_1^2} + \f{1}{f_2^2} ~=~ \f{1}{v^2}\,,
 \end{equation}
 which imply
 \begin{equation}
  \gz ~=~ g\sqrt{1+x^2}    \,,  \qquad\qquad
  \gb ~=~ g\sqrt{1+x^{-2}} \,,
  \end{equation}
 where $g$ is the gauge coupling of the SM gauge group $SU(2)_W$ and
 the vacuum expectation value $v$ is determined by the Fermi constant
 $\,v=(\sqrt{2}G_F)^{-1/2}\,$.\,  The model parameter space is such
 that $\,x^2\ll 1\,$ and $\,y^2={\cal O}(1)\,$.\,  
 
 We may build up the gauge boson mass matrix, $\MMW$, and its dual, $\MMWT$, in the 3-site model by starting from the matrices,
  \begin{equation}
\CF = \frac{1}{2} \left\lgroup
\begin{array}{cc}
f_1 & 0 \\
0 & f_2
\end{array}
\right\rgroup\,, \qquad\qquad 
\CD_W = \left\lgroup
\begin{array}{cc}
1 & -1 \\
0 & 1
\end{array}\right\rgroup\,,\qquad\qquad 
\CG = \left\lgroup
\begin{array}{cc}
g_0 & 0 \\
0 & g_1
\end{array}\right\rgroup\,,
 \end{equation}
 multiplying them, as in Eq. (\ref{eq:Q}) to form,
 \begin{equation}
\QQB = \frac{1}{2} \left\lgroup
\begin{array}{cc}
g_0 f_1 & - g_1 f_1 \\
0 & g_1 f_2
\end{array}
\right\rgroup\,,
 \end{equation}
 and then using Eq. (\ref{eq:mwmwdq}) to arrive at
  \beqa
  &&
  \dis\MMW ~=~\f{1}{4}
  \left\lgroup
  \ba{cc}
  \dis\gzz f_1^2     & -\dis\gz\gb f_1^2
  \\[3mm]
  -\dis\gz\gb f_1^2  & \dis\gbb (f_1^2+f_2^2)
  \ea
  \right\rgroup
  ~=~ \dis\(\!\f{g_1f_1}{2}\!\)^{\!2}
  \left\lgroup
  \ba{cc}
  x^2 & -x
  \\[3mm]
  -x    &  1+y^2
  \ea
  \right\rgroup ,
  \\[4mm]
  &&
 \dis\MMWT ~=~\f{1}{4}
  \left\lgroup
  \ba{cc}
  \dis (\gzz +\gbb ) f_1^2     & -\dis\gbb f_1f_2
  \\[3mm]
  -\dis\gbb f_1f_2          & \dis\gbb f_2^2
  \ea
  \right\rgroup
  ~=~ \dis\(\!\f{g_1f_1}{2}\!\)^{\!2}
  \left\lgroup
  \ba{cc}
  1+x^2 & -y
  \\[3mm]
  -y    &  y^2
  \ea
  \right\rgroup .
  \eeqa
  For the typical case of $\,x^2\ll 1\,$ and $\,y=1\,$ (i.e., $f_1=f_2=\sqrt{2}v$\,),
  the mass eigenvalues of $W_0$ and $W_1$ gauge bosons can be expanded as
  \beqa
  M_{W0}^2 &=& \(\f{\gz v}{2}\)^2\(1-\f{x^2}{4}+\f{x^6}{64}+\cdots\)
  \,,
  \\[2mm]
  M_{W1}^2 &=& \(g_1v\)^2\(1+\f{x^2}{4}+\f{x^4}{16}+\cdots\)
  \,,
  \eeqa
  and the corresponding rotation matrices for diagonalizing $\MMW$ and $\MMWT$ are,
  up to ${\cal O}(x^5)$,
  \beqa
  \RRB  &\!\!=\!\!& \left\lgroup
  \ba{ll}
  1-\f{x^2}{8}-\f{5x^4}{128} &  -\f{x}{2}-\f{x^3}{16}+\f{9x^5}{256}
  \\[3mm]
  \f{x}{2}+\f{x^3}{16}-\f{9x^5}{256}  &
  1-\f{x^2}{8}-\f{5x^4}{128}
  \ea
  \right\rgroup  \equiv
  \left\lgroup
  \ba{ll}
  a & b
  \\[3mm]
c  & d
  \ea
  \right\rgroup  ,
  \\[4mm]
  \RRBT &\!\!=\!\!& \left\lgroup
  \ba{ll}
  \f{1}{\sqrt{2}}\(1-\f{x^2}{4}-\f{x^4}{32}\)
  &
  -\f{1}{\sqrt{2}}
  \(1+\f{x^2}{4}-\f{x^4}{32}\)
  \\[3mm]
  \f{1}{\sqrt{2}}
  \(1+\f{x^2}{4}-\f{x^4}{32}\)
  &
  \f{1}{\sqrt{2}}\(1-\f{x^2}{4}-\f{x^4}{32}\)
  \ea
  \right\rgroup 
   \equiv
  \left\lgroup
  \ba{ll}
  \tilde{a} & \tilde{b}
  \\[3mm]
\tilde{c}  & \tilde{d}
  \ea
  \right\rgroup  \,.
  \eeqa
  It is clear that these matrices are orthogonal, as needed to satisfy the completeness relations for the gauge and Nambu-Goldstone KK wavefunctions.  One may also readily verify that $\RRB$ and $\RRBT$ do work together to diagonalize $\QQB$ as in Eq. (\ref{eq:qrrtdiag}).  The schematic forms of the rotation matrices at the far right will be used below to illustrate the derivation of the couplings.
  
\subsection{Couplings}
  
  Having obtained the matrices $\RRB$ and $\RRBT$, we may now calculate the couplings among gauge bosons and Nambu-Goldstone bosons.  Here, we quote those that are relevant for studying $nn \to mm$ two-body scattering processes.  In deriving explicit formulae for the couplings from the general equations in Section \ref{sec:higgsless}, it is important to keep in mind that the indices on the $\RRB$ matrices refer to sites along the linear moose, while the indices on the $\RRBT$ matrices refer to links; this is especially important for the couplings between gauge and Nambu-Goldstone states because the left-most site is labeled "0" while the left-most link is labeled "1.  We give a schematic form for the first coupling of each category to show how it relates to entries in the $\RRB$ and $\RRBT$ matrices.
  
  First, we give the minimal independent set (recalling Bose symmetry) of quartic gauge couplings, as obtained from Eq. (\ref{eq:def4}),
  \beqa
  G_4^{0000}  &=&  g^2\[1+\f{9}{16}x^2-\f{15}{32}x^4+\f{13}{256}x^6+ \cdots \] = g_0^2 a^4 + g_1^2 c^4,
  \nonumber\\
  G_4^{0011}&=& \f{g^2}{4}\[1+2x^2+\f{3}{4}x^4-\f{x^6}{2}+ \cdots \] ,
  \\
  G_4^{1111} &=& \f{g^2}{x^2}\[1+\f{1}{2}x^2-\f{9}{16}x^4+\f{3}{32}x^6
  +\f{51}{256}x^8 + \cdots \] \,,
  \nonumber
  \eeqa  
  and then the triple gauge couplings from Eq. (\ref{eq:def3})
    \beqa
  G_3^{000} &=& g\[1+\f{1}{4}x^2-\f{35}{128}x^4+\f{31}{256}x^6+ \cdots\]  = g_0 a^3 + g_1 c^3,
  \nonumber\\
  G_3^{001} &=& -\f{g}{4}x\[1+\f{1}{8}x^2-\f{57}{128}x^4+\f{129}{1024}x^6+ \cdots\] ,
  \nonumber\\
  \\[-5mm]
  G_3^{011} &=& \f{g}{2}\[1+\f{7}{8}x^2-\f{5}{128}x^4 -\f{109}{1024}x^6 + \cdots \] ,
  \nonumber\\
  G_3^{111}&=& \f{g}{x}\[1+\f{1}{8}x^2-\f{65}{128}x^4 +\f{25}{1024}x^6 + \cdots \] 
  \,.
  \nonumber
  \label{eq:ggg3}
  \eeqa
  Next, there are the  quartic couplings among Nambu-Goldstone bosons, from Eq. (\ref{eq:deft4})
  \beq
  \label{eq:Gpi4}
  \ba{ll}
  \Gt_4^{0000}= \Gt_4^{1111}&=~\dis
   \f{1}{4}\[1+\f{1}{4}x^4-\f{1}{16}x^8+\f{1}{64}x^{12}+ \cdots \] = \frac{v^2}{f_1^2} \tilde{a}^4 + \frac{v^2}{f_2^2} \tilde{c}^4 \,, \\ [4mm]
  \Gt_4^{0011}= \Gt_4^{1100}&=~\dis
   \f{1}{4}\[1-\f{1}{4}x^4+\f{1}{16}x^8-\f{1}{64}x^{12}+ \cdots \],
  \ea
  \eeq
  the $\pi\pi\pi V$ quartic couplings from Eq. (\ref{eq:deft41})
 \begin{eqnarray}
 \Gt_{41}^{0000} & = & - \frac{g}{4} \[ 1 + \frac{3}{8} x^2 + \frac{7}{128} x^4 + \frac{179}{1024} x^6 + \cdots\] =   \frac{v}{f_1} \tilde{a}^3 \(g_1 c - g_0 a \) +  \frac{v}{f_2} \tilde{c}^3 \( - g_1 c\) \nonumber \,,\\
 \Gt_{41}^{0011} & = & -\frac{g}{2x} \[ 1 + \frac{5}{8} x^2 - \frac{37}{128} x^4 - \frac{103}{1024} x^6 + \cdots\] \,,\nonumber \\
   \Gt_{41}^{1100} & = & -\frac{g}{4} \[ 1 + \frac{3}{8} x^2 - \frac{57}{128} x^4 - \frac{13}{1024} x^6 + \cdots\] \,,\\
 \Gt_{41}^{1111} & = &- \frac{g}{2x} \[ 1 + \frac{5}{8} x^2 + \frac{27}{128} x^4 + \frac{217}{1024} x^6 + \cdots\] \,,\nonumber
  \end{eqnarray}
  and the $\pi\pi VV$ quartic couplings from Eq. (\ref{eq:deft42})
 \begin{eqnarray}
\Gt_{42}^{0000} &=& - \Gt_{42}^{0011} = - \frac{g^2}{4} \[1 + \frac{1}{2}x^2 - \frac{5}{8} x^4  + \cdots \]  = -g_0 g_1 \tilde{a}^2 a c \,,\nonumber \\
\Gt_{42}^{1111} &=& - \Gt_{42}^{1100} = - \frac{g^2}{4} \[1 + \frac{3}{2} x^2 + \frac{3}{8}x^4 - \frac{1}{4} x^6 + \cdots \]\, .
  \end{eqnarray}
  Finally, there are the $\pi\pi V$ couplings from Eq. (\ref{eq:deft3})
  \beq
  \label{eq:Gpi3}
  \ba{ll}
  \Gt_3^{000} &=~\dis g
   \[1+\f{1}{4}x^2-\f{35}{128}x^4+\f{31}{256}x^6+ \cdots \] = \tilde{a}^2 \( g_1 c + g_0 a\) + \tilde{c}^2 \(g_1 c\)  \, ,
  \\[4mm]
  \Gt_3^{001} &=~\dis \f{g}{x}
   \[1+\f{1}{8}x^2-\f{33}{128}x^4+\f{185}{1024}x^6+ \cdots \],
   \\[4mm]
\Gt_3^{110} &=~\dis g\left[ 1+\frac{3}{4}x^2-\frac{11}{128}x^4-\frac{7}{128}x^6+\cdots
\right],
\\[4mm]
  \Gt_3^{011} &=~
  \dis \frac{g x}{4}\[ 1+\f{5}{8}x^2 - \f{33}{128}x^4 - \f{83}{1024}x^6+ \cdots \],
  \\[4mm]
  \Gt_3^{111} &=~
  \dis \f{g}{x}\[ 1+\f{1}{8}x^2 - \f{65}{128}x^4 +
   \f{25}{1024}x^6+ \cdots \].
  \ea
  \eeq
  We omit the $\pi VV$ couplings since they do not affect the processes of interest.

\subsection{Identities}

We have verified that the ET sum rules (and equivalent completion relations or Ward-Takahasi identities) are satisfied by the couplings and masses given above.  Here, we give a few straightforward examples.

As mentioned above, the manifest orthogonality of the matrices $\bbR$ and $\tilde{\bbR}$ shows that the completeness relations are satisfied.  Equivalently, one can show that the first two sum rules  in Table \ref{tab:sumrules} (Eqs. (\ref{eq:E4-sum1}) and (\ref{eq:E4-sum2})) hold: 
 \beq
 \ba{ll}
 \dis
 G_4^{0000} &=~ \dis\sum_{k=0}^1 (G_3^{00k})^2 ~=~
 g^2\[1+\f{9}{16}x^2 -\f{15}{32}x^4+\f{13}{256}x^6 + {\cal O}(x^7)\] ,
 \\[6mm]
 G_4^{0011} & =~ \dis\sum_{k=0}^1 (G_3^{01k})^2 ~=~
 \sum_{k=0}^1 G_3^{00k} G_3^{11k}
 ~=~ \dis\f{g^2}{4}\[1+2x^2 +\f{3}{4}x^4-\f{1}{2}x^6 + {\cal O}(x^7)\] ,
 \\[6mm]
 G_4^{1111} &=~ \dis\sum_{k=0}^1 (G_3^{11k})^2 ~=~
 \dis\f{g^2}{x^2}\[1+\f{1}{2}x^2-\f{9}{16}x^4+\f{3}{32}x^6
  +\f{51}{256}x^8\] + {\cal O}(x^9)\,.
 \ea
 \eeq

It is straightforward to show that the Ward-Takahashi Identify for $\tilde{G}_3$ (Eq. \ref{eq:WT-t3}) is satisfied in the three-site model.  For instance,
it predicts the four equalities
\begin{equation}
\tilde{G}_3^{000} = G_3^{000}\,, \qquad \qquad \tilde{G}_3^{111} = G_3^{111} \,,\qquad \qquad \tilde{G}_3^{001} M_0^2 = G_3^{001} (2 M_0^2 - M_1^2)\,, \qquad\qquad \tilde{G}_3^{011} M_0 M_1 = G_3^{011} M_0^2\,.
\end{equation}
The first two may be verified by inspection of Eqs. (\ref{eq:ggg3}) and (\ref{eq:Gpi3}); for the others, we find,
\begin{eqnarray}
\tilde{G}_3^{001} M_0^2 &=& G_3^{001} (2 M_0^2 - M_1^2) = \frac{g^3 v^2}{4 x} \( 1 + \frac{7}{8} x^2 - \frac{53}{128} x^4 - \frac{29}{1024} x^6 + \cdots \)\,, \nonumber \\
\tilde{G}_3^{011} M_1 &=& G_3^{011} M_0 = \frac{g^2 v}{4} \( 1 + \frac{5}{4} x^2 + \frac{3}{32} x^4 - \frac{27}{128} x^6 \)\,.
\end{eqnarray}
Knowing that this WTI and the NG-boson completeness relation are verified confirms that their equivalents, the sum rules in Eqs. (\ref{eq:E0-sum1}), (\ref{eq:E0-sum2}), (\ref{eq:SR-E0-LLTT-1in}), and (\ref{eq:SR-E0-LLTT-2in}),  hold in the three-site model.

The sum rule of Eq. (\ref{eq:SR-E0-LLTT-3in}), or equivalently, the deconstruction Identity for $\tilde{G}_{42}$, Eq. (\ref{eq:WT-t42}), leads to the following equalities in the three-site model,
\begin{eqnarray}
\tilde{G}_{42}^{0000} M_0^2 &=& G_4^{0000} M_0^2 - \( G_3^{000}\)^2 M_0^2 - \( G_3^{001} \)^2 M_1^2 = - \frac{g^4 v^2}{16} \( 1 + \frac{5}{4} x^2 - \frac{1}{2} x^4 - \frac{37}{64} x^6 + \cdots \) \,,
\nonumber \\
\tilde{G}_{42}^{1111} M_1^2 &=& G_4^{1111} M_1^2 - \( G_3^{110}\)^2 M_0^2 - \( G_3^{111} \)^2 M_1^2 = - \frac{g^4 v^2}{4 x^2} \( 1 + \frac{11}{4} x^2 + \frac{41}{16} x^4 + \frac{3}{4} x^6 + \cdots \) \,,
\end{eqnarray}
which we have verified explicitly.

 Finally, the relationship
 (\ref{eq:G41a}) between the $\pi\pi\pi\pi$ and $\pi\pi\pi V$ couplings implies the following equalities in the three-site model,
 \beq
  \ba{l}
 \f{v}{2M_0}\Gt_{41}^{0000} ~=~   \f{v}{2M_1}\Gt_{41}^{1111} 
     ~=~ -\Gt_4^{0000}~=~ -\Gt_4^{1111} \,,
 \\ [7mm]
  \f{v}{2M_1}\Gt_{41}^{0011} ~=~\f{v}{2M_0}\Gt_{41}^{1100} 
      ~=~  -\Gt_4^{0011} ~=~ -\Gt_4^{1100} \,.
 \ea
\eeq
and we have verified both. These also follow directly from combining the deconstruction identities (\ref{eq:WT-t4}) and (\ref{eq:WT-t41}).


\end{document}